\documentclass[aps,pra,reprint,amsmath,amssymb,showpacs,superscriptaddress,floatfix]{revtex4-1}

\usepackage[colorlinks,citecolor=blue,linkcolor=blue]{hyperref}
\usepackage{graphicx}
\usepackage{dcolumn}
\usepackage{bm}
\usepackage{color}
\usepackage{txfonts}

\newcommand{\IEF}{Institut d'Electronique Fondamentale, CNRS, Univ. Paris-Sud, Universit\'e Paris-Saclay, 91405 Orsay, France}
\newcommand{\LSPM}{Laboratoire des Sciences des Proc{\'e}d{\'e}s et des Mat{\'e}riaux, CNRS-UPR 3407, Universit{\'e} Paris 13, Sorbonne Paris Cit{\'e}, 93430 Villetaneuse, France}
\newcommand{\ITMO}{International laboratory "MultiferrLab", ITMO University, St. Petersburg, Russia}
\newcommand{\NIMS}{National Institute for Materials Science, Tsukuba 305-0047, Japan}

\begin{document}
\title{Probing the Dzyaloshinskii-Moriya interaction in CoFeB ultrathin films using domain wall creep and Brillouin light spectroscopy}
\author{R. Soucaille}
\email{remy.soucaille@u-psud.fr}
\affiliation{\IEF}
\author{M. Belmeguenai}
\affiliation{\LSPM}
\author{J. Torrejon}
\affiliation{\NIMS}
\author{J.-V. Kim}
\author{T. Devolder}
\affiliation{\IEF}
\author{Y. Roussign\'e}
\author{S.-M. Ch\'erif}
\affiliation{\LSPM}
\author{A. A. Stashkevich}
\affiliation{\LSPM}
\affiliation{\ITMO}
\author{M. Hayashi}
\affiliation{\NIMS}
\author{J.-P. Adam}
\email{jean-paul.adam@u-psud.fr}
\affiliation{\IEF}

\date{\today}

\begin{abstract}
We have characterized the strength of the interfacial Dyzaloshinskii-Moriya interaction (DMI) in ultrathin perpendicularly magnetized CoFeB/MgO films, grown on different underlayers of W, TaN, and Hf, using two experimental methods. First, we determined the effective DMI field from measurements of field-driven domain wall motion in the creep regime, where applied in-plane magnetic fields induce an anisotropy in the wall propagation that is correlated with the DMI strength. Second, Brillouin light spectroscopy was employed to quantify the frequency non-reciprocity of spin waves in the CoFeB layers, which yielded an independent measurement of the DMI. By combining these results, we show that DMI estimates from the different techniques only yield qualitative agreement, which suggests that open questions remain on the underlying models used to interpret these results.
\end{abstract}

\maketitle

\section{Introduction}

Magnetic order in ultrathin films is largely driven by surface and interface effects that can lead to the appearance of new energy terms which are not present in the bulk. A well-known example is perpendicular magnetic anisotropy, which describes an easy anisotropy axis that appears perpendicular to the film plane in ultrathin films and multilayers. This phenomenon is induced by interface-driven changes to the orbitals in the ferromagnet and is important for current magnetic storage technologies because higher bit-densities can be achieved with perpendicular media. A more recent example of present interest concerns chiral interactions of the Dzyaloshinskii-Moriya form, which appear in similar ultrathin films in contact with a normal metal possessing large spin-orbit coupling. Despite its prediction over two decades ago~\cite{Fert:1980, fert_magnetic_1990}, compelling experimental evidence of its importance in ultrathin film systems has only been obtained during the past few years. For example, it has been shown that this interfacial Dzyaloshinskii-Moriya interaction (DMI) leads to spin spirals in Mn monolayers on W(110)~\cite{Bode:2007}, nanoscale skyrmion lattices in Fe monolayers on Ir(111)~\cite{heinze2011spontaneous}, and isolated skyrmions in Pd/Fe bilayers on Ir (111)~\cite{Romming:2013}.

While the DMI strength in such epitaxially-grown monolayer systems correlates well with predictions from \emph{ab initio} calculations~\cite{heinze2011spontaneous}, the situation for polycrystalline films grown by sputtering is not as clear. Recent experiments have shown that the interfacial DMI can be sufficiently large to promote chiral spin states in systems based on Pt/Co, where room temperature skyrmions have been reported in Pt/Co/Ir multilayers~\cite{Moreau_luchaireCNNano2016}, Pt/Co/MgO films~\cite{Boulle:2016} and in Pt/Co/Ta~\cite{Woo:2016} films, and homochiral N{\'e}el walls have been observed in Pt/Co/AlOx~\cite{Tetienne2015, Benitez:2015}. The results appear to be consistent with measurements of the frequency non-reciprocity of spin wave propagation with Brillouin light spectroscopy~\cite{Belmeguenai:2015} and predictions from electronic structure calculations~\cite{Freimuth:2014}. However, for ferromagnetic alloys such as CoFe or CoFeB, consensus is yet to be reached on the strength of possible induced chiral interactions at ferromagnet/heavy metal interfaces. Indeed, experiments have shown that chiral magnetic bubbles can be nucleated in Ta/CoFeB/TaOx~\cite{Jiang:2015}, yet other studies on the similar Ta/CoFeB/MgO system using single spin magnetometry have shown no evidence of any chiral interaction present~\cite{Tetienne2015}.

In this article, we seek to clarify the issue of the DMI strength in perpendicularly magnetized CoFeB/MgO films deposited on different heavy-metal underlayers, namely Hf, TaN, and W. In previous work, current-driven domain wall motion under applied magnetic fields was used to estimate the DMI strength for different thicknesses of these underlayers~\cite{Natcomtorrejon2014}. Here, we revisit this problem by employing techniques that do not rely on current-dependent spin torques, which possess different components (adiabatic, non-adiabatic, spin-Hall-like and Rashba-like fields) and whose collective effect on the motion of domain walls remains unclear. Instead, we employ two techniques to probe the Dzyaloshinskii-Moriya energy without strong assumptions on the dynamics. First, we use field-driven domain wall motion in the creep regime in which wall velocities are governed by a power law, where the dominant term arises from changes in the domain wall (elastic energy) due to the DMI~\cite{Je:2013, Hrabec:2014}. Second, we use Brillouin light spectroscopy to measure the non-reciprocal propagation of spin waves in the Damon-Eshbach geometry, where it has been shown that the frequency non-reciprocity is a direct measure of the DMI constant~\cite{Moon:2013, Di:2015, Belmeguenai:2015}. Both methods have been employed on the same multilayer films. While qualitative agreement is found for most cases, numerical estimates of the DMI strength for a given underlayer can differ considerably, which suggests inconsistencies remain in the underlying assumptions used to interpret these data.

This article is organized as follows. In Section II, we describe the sample  structure and deposition methods. In Section III, we present results from field driven domain wall motion in the creep regime, which was characterized using Kerr effect microscopy. In Section IV, we present results from Brillouin light spectroscopy measurements in which frequency nonreciprocity is probed for spin waves in the Damon-Eshbach geometry. An analysis of the DMI strength obtained using the two methods is presented in Section V and some concluding remarks are given in Section VI.
%
%
%
%
%
%
\section{Samples}
Our multilayers were grown by sputtering and have the following nominal structure: Si/$\mathrm{SiO_2}$/X($t$)/$\mathrm{Co_{20}Fe_{60}B_{20}}$(1)/MgO(2)/Ta(1), where figures in parentheses denote the film thicknesses in nanometers. They were annealed in vacuum at 300 $^\circ$C for 1 hour. We consider in this study four different underlayers X: W (2 nm), W(3 nm), TaN (1 nm) and Hf (1 nm). The structural properties of the samples with metallic buffers were studied in \cite{LiuHayashiAPL2015}. High resolution transmission electron microscopy showed that W, Ta and Hf underlayers were amorphous in the thickness range investigated here. Larger thicknesses --i.e. 2 nm for Hf, and 3 to 5 for W-- were required to obtain crystalline transitions within the underlayer. The TaN buffers are also essentially amorphous~\cite{Natcomtorrejon2014}, while lattice fringes in MgO layers indicate a textured crystalline character that is partially replicated in the CoFeB layer \cite{LiuHayashiAPL2015}.
%
%
%
%
%
%
\section{Domain wall motion in the creep regime}

\begin{figure}
	\centering
	\includegraphics[width=8.5cm]{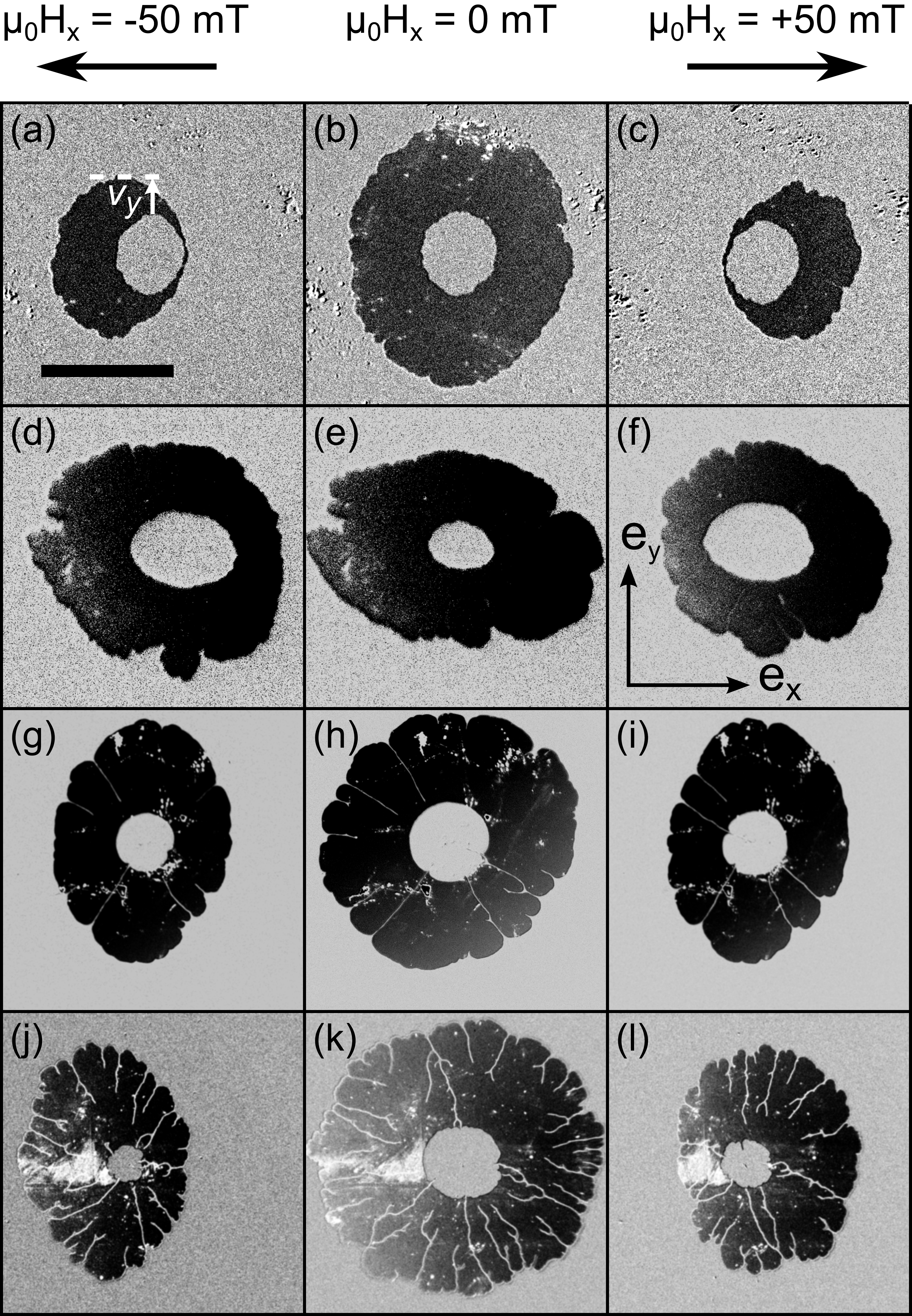}
		\caption{\label{Fig_Kerr}Differential Kerr images illustrating the expansion of a nucleated domain, where the black region indicates the area swept by the domain wall during the field pulse. The images compare motion under a pulsed perpendicular field, $H_z$, with an additional in-plane static field, $H_x$. The scale bar indicated in (a) represents 125~$\mu$m for (a)-(f) and 250~$\mu$m for (g)-(l).
(a)-(c) W (2 nm) underlayer under $\mu_0 H_z = 2.8$ mT with a pulse duration of 5~s in (a,c) and 50~s in (b).
(d)-(f) W (3 nm) underlayer under $\mu_0 H_z = 2.9$ mT with a pulse duration of 1.5~s in (d, f) and 15~s in (b).
(g)-(i) TaN (1 nm) underlayer under $\mu_0 H_z = 0.83$ mT with a pulse duration of 0.4~s in (g, i) and 1.5~s in (h).
(j)-(l) Hf (1 nm) underlayer under $\mu_0 H_z = 2.9$ mT with a pulse duration of 75~ms in (j, l) and 250~ms in (b).
The in-plane field is $\mu_0 H_x = -50$~mT in (a, d, g, j), $\mu_0 H_x = 0$~mT in (b, e, h, k), and $\mu_0 H_x = +50$~mT in (c, f, i, l).
}
\end{figure}

Domain wall motion in our ultrathin CoFeB layers was studied with a magneto-optical polar Kerr effect microscope with vector field capability. To determine the domain wall velocity, we first nucleated an approximately circular domain at the center of the field of view of the microscope. We then applied pulses of the perpendicular magnetic field, down to 15 $\mu$s in duration, and determined the distance travelled by the domain walls for both domain expansion and compression of the nucleated domain. The velocity was then estimated by dividing the total distance traveled by the domain wall during the duration of the field pulse~\cite{Metaxas2007}. These experiments were conducted in the presence of an additional static in-plane applied field, which was varied between -150 and 150 mT. An example of such measurements is given in Fig.~ \ref{Fig_Kerr}.
We paid particular attention to the placement of the sample with respect to the electromagnets in order to minimize artifacts due to crosstalk between in-plane and perpendicular field. This involved a specific procedure to precisely align the electromagnet in the sample plane which consisted in examining domain expansion for both polarities of the nucleated domain and applied in-plane fields.

\begin{figure}
	\centering\includegraphics[width=8.5cm]{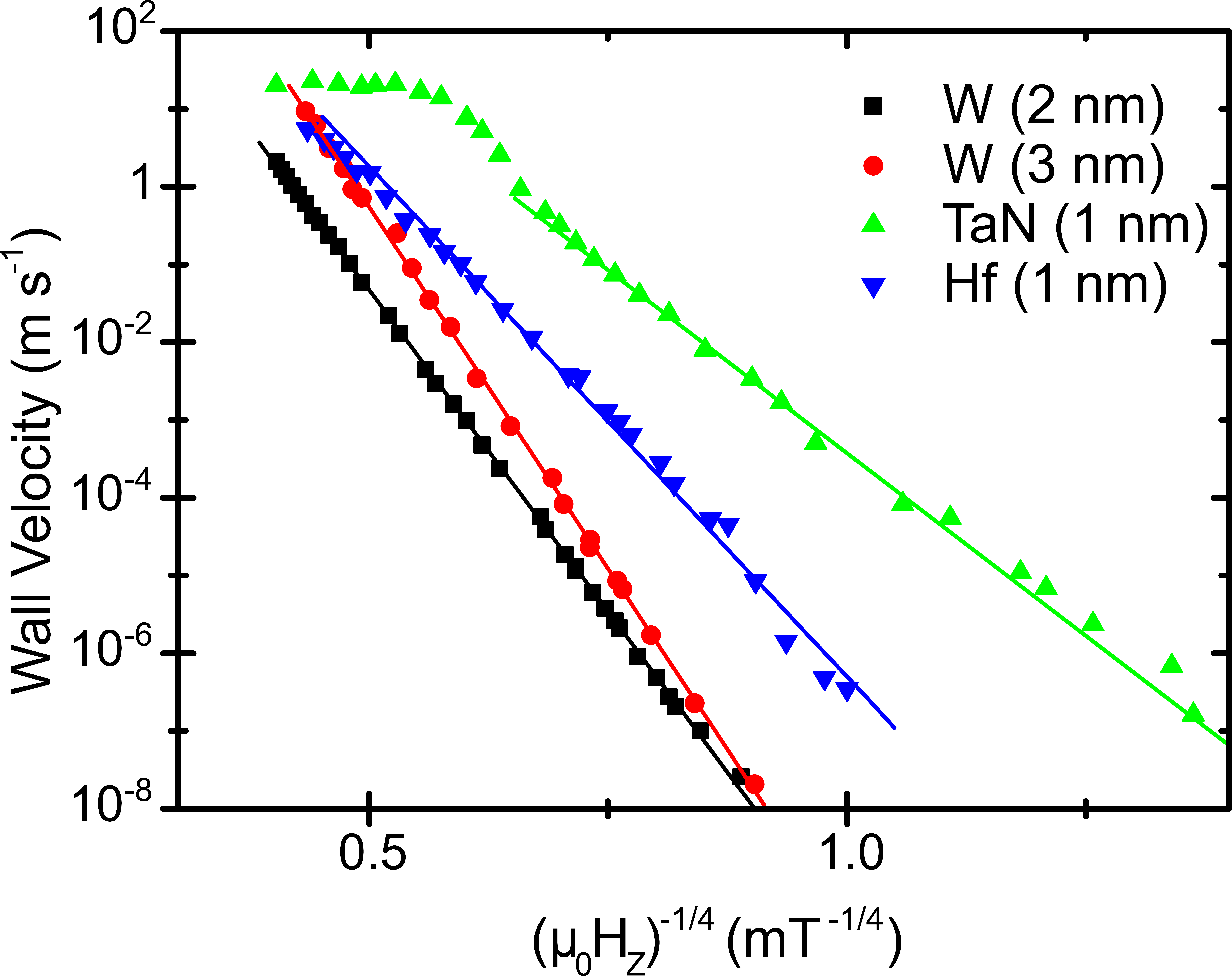}
	\caption{(Color online) Domain wall velocity $v$ in CoFeB as a function of perpendicular field $\mu_0 H_z$ for different underlayers in zero in-plane field. The solid lines corresponds to fits based on the creep model [Eq.~(\ref{eq:creep})].
	\label{Fig_creep}}
\end{figure}

For perpendicular fields well below the depinning threshold, the wall motion in ultrathin PMA films is described by the creep model. In this regime, the wall dynamics can be linked to the motion of a one-dimensional elastic string in a two dimensional disordered potential, where motion is driven by thermal activation and involves a series of avalanches. It is well established that the dependence of the wall velocity, $v$, on the applied perpendicular field, $H_z$, in this regime can be described by the following Arrhenius-type relation~\cite{Lemerle:1998},
\begin{equation}
v(H_z) = v_0 \exp\left[- \alpha\left(H_z\right)^{-\frac{1}{4}} \right],
\label{eq:creep}
\end{equation}
where $v_0$ is a velocity prefactor and $\alpha$ is a function of $H_z$ that depends on the wall (elastic energy), the pinning potential, and the thermal energy $k_B T$. A plot of the measured domain wall velocity for the different underlayers is given in Fig.~\ref{Fig_creep}. 
By using a log-linear scale and by plotting the measured velocities as a function of $H_z^{-1/4}$, it is easy to identify the range of fields over which the wall motion remains in the creep regime. For the TaN underlayer, we observe a deviation from the creep behavior at higher applied fields where a change in the linear variation in Fig.~\ref{Fig_creep} can be seen, but for all other samples the motion remains in the creep regime for the range of applied fields considered. For the W underlayers, we note that the underlayer thickness plays an important role on the wall velocity (for the same nominal CoFeB film), where a clear difference in the slope of the velocity curves can be seen. This suggests that the domain wall pinning potential in the two multilayer stacks is different, which might arise from a difference in film morphology

\begin{figure}
\centering\includegraphics[width=8.5cm]{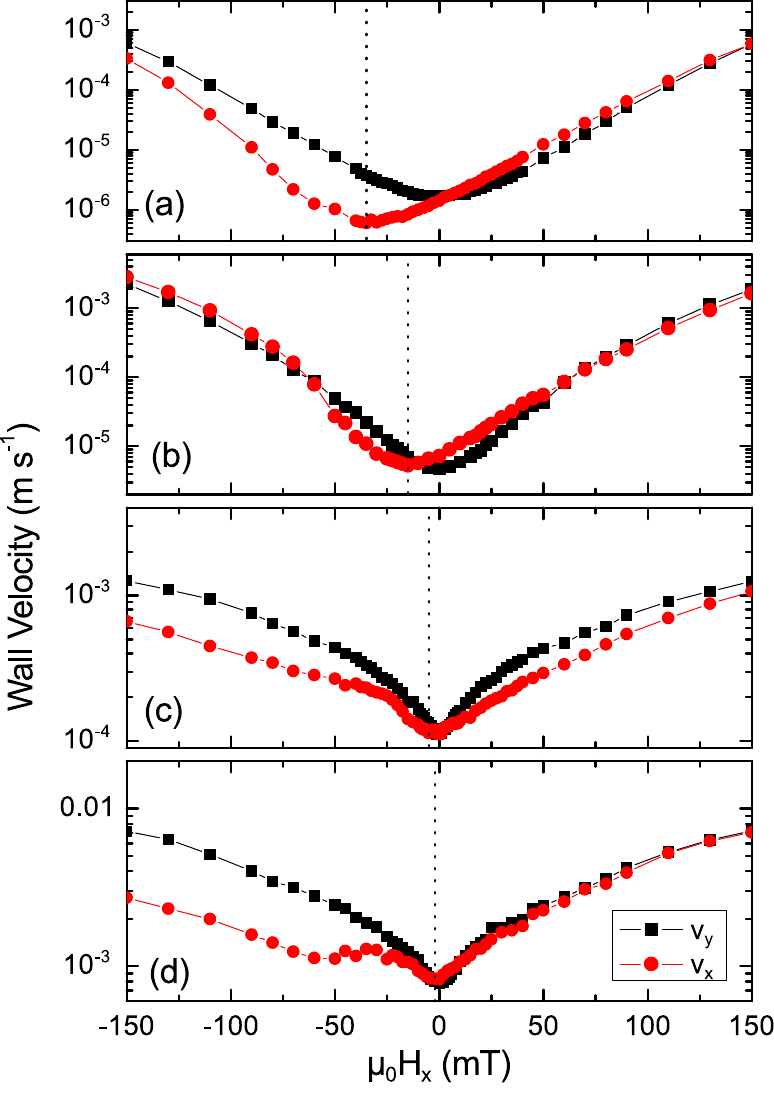}
\caption{(Color online) Domain wall velocity as a function of in-plane applied field $H_x$ for propagation along (squares) and perpendicular (circles) to the field direction. The propagation takes place under a static perpendicular field, $H_z$. (a) 2~nm thick W underlayer at $\mu_0 Hz = 2.8$ mT. (b) 3~nm thick W underlayer at $\mu_0 Hz = 2.9$ mT. (c) 1~nm thick TaN underlayer at $\mu_0 Hz = 0.83$ mT. (d) 1~nm thick Hf underlayer at $\mu_0 Hz = 2.9$ mT. The dashed vertical line indicates the offset field, $H_{\rm offset}$.
\label{Fig_Hip}}
\end{figure}

To investigate the presence of a Dzyaloshinskii-Moriya interaction, we measured the wall motion in the presence of a finite in-plane magnetic field. As previous studies have shown~\cite{Kabanov:2010, Je:2013, Hrabec:2014, Lavrijsen:2015, Vanatka:2015}, an in-plane magnetic field can break the cylindrical symmetry of the domain wall energy, which is key to revealing the presence of a chiral interaction that prefers a given handedness for the domain wall. This asymmetry can be seen in the domain expansion in Fig.~\ref{Fig_Kerr}, particularly for the W underlayers [Fig.~\ref{Fig_Kerr}(a)-(c)], where the propagation is very different along the axis of the applied in-plane magnetic field. For the TaN [Fig.~\ref{Fig_Kerr}(d)-(f)] and Hf (not shown) underlayers, this asymmetry along the applied field direction is less pronounced but other features, such as an elliptical shaped bubble, are seen instead. In order to quantify this asymmetry, we plot in Fig.~\ref{Fig_Hip} the domain wall velocity as a function of the in-plane applied field $H_x$ at constant $H_z$, where the velocity along the field axis, $v_x$, and perpendicular to the field axis, $v_y$, for the different underlayers is shown.
For each sample, a different perpendicular field $H_z$ was used to keep the wall velocities within a similar range to facilitate comparison. For the range of in-plane fields studied, the $V_y(H_x)$ curve is found to be symmetric with respect to $H_x = 0$ for all underlayers, although the curvature of the apex of the `V'-shaped curve is found to vary with the underlayer. For instance, the variation is found to be sharp for the TaN underlayer [Fig.~\ref{Fig_Hip}(c)], while the transition is smoother for both W thicknesses [Fig.~\ref{Fig_Hip} (a)-(b)]. The curves have been obtained from domain expansion for both up and down domains. We also note that the bubble growth has a strong asymmetric character in the direction parallel to the in-plane field axis  [Fig.~\ref{Fig_Hip}(a)], which leads to difficulties in defining accurately the wall velocity in the direction perpendicular to the in-plane field axis ($\mathrm{v_y}$). Our convention is to define it from the vertical distance between the two points where the DW is parallel to the in-plane field [see arrow in Fig.~\ref{Fig_Kerr}(a)].

For wall motion along the field axis, on the other hand, the $v_x(H_x)$ curves exhibit a deformed `V'-shaped curve and are displaced along the field axis. The magnitude of this displacement differs for each underlayer. We determined the value of the applied in-plane field that results in a minimum in the DW velocity for each underlayer, which is denoted as $H_{\rm offset}$ in Table~\ref{Tab_Hoffest}. We find that the measured offset fields have the same sign for all underlayers (W, Hf and TaN), where the shift is towards negative values of the in-plane field. This asymmetry can be understood as follows. The in-plane field favors a certain orientation of the magnetization within the domain wall. If an ``up'' bubble domain is nucleated (i.e., $M_z > 0$ within the nucleated bubble), then a positive in-plane field $H_x >0$ favors a right-handed domain wall on the right side of the bubble, while a left-handed domain wall is favored on the left side of the bubble when viewed from above. In the absence of a chiral interaction, both handedness are degenerate and a circular expansion is expected. However, the presence of a DMI breaks this symmetry by favoring one handedness over the other, which means that the wall energies associated with motion parallel and antiparallel to the applied field are no longer equivalent. The minimum in the velocity curve coincides with the applied field that compensates the internal chiral DMI field, which therefore allows its magnitude to be estimated. Similarly, motion perpendicular to the applied field axis remains symmetric with respect to the applied field because the wall energies remain degenerate along this direction. We will revisit this analysis in more detail in Sec. V.
%
%
%
%
%
%
\section{Brillouin light spectroscopy}
In addition to domain wall motion, we have also characterized the DMI using Brillouin light scattering (BLS) measurements on the same multilayer films. In contrast to domain wall creep, where the domain wall energy is the quantity affected by the DMI, the BLS measurements probe propagating spin waves in the ultrathin CoFeB film where the DMI manifests itself as a propagation nonreciprocity for the spin waves, i.e., for a given wavelength, the two spin waves propagating in opposite directions have different frequencies when the static magnetization is in-plane and the wave vector is perpendicular to the static magnetization. For the interfacial form of the DMI present in ultrathin ferromagnets in contact with a heavy metal underlayer, spin wave propagation remains reciprocal in the absence of any in-plane applied fields in the uniform state, far from boundary edges~\cite{Garcia-Sanchez:2014}. However, when the magnetization is tilted away from its equilibrium orientation along the easy axis (perpendicular to the film plane) by an applied field, a nonreciprocity appears for spin wave propagating perpendicular to the magnetization direction. When the magnetization is saturated in the film plane by the applied field, the geometry corresponds to the Damon-Eshbach geometry for in-plane magnetized films in which magnetostatic surface spin waves exhibit nonreciprocal propagation in the direction perpendicular to the magnetization. For our CoFeB layers, we argue that the nonreciprocity is driven primarily by the presence of the DMI rather than dipolar effects~\cite{Moon:2013}.

In our BLS experiment, we applied in-plane magnetic fields that were sufficiently large to saturate the magnetization in the film plane. We employed the backscattering geometry and a 2$\times$3 pass Fabry-Perot interferometer to investigate propagating spin waves in the Damon-Eshbach geometry for different values of the incident wave vector, by measuring the frequency shifts of the inelastically scattered light with respect to the frequency of the incident laser beam with a wavelength of $\lambda$ = 532 nm. For each wave vector, the spectra were obtained after accumulating photons for a few hours in order to determine the scattered line position to an accuracy better than 0.1~GHz. The Stokes (S) and anti-Stokes (AS) frequencies were then determined from Lorentzian fits to the BLS spectra in order to obtain the desired frequency difference, $ \Delta f = f_S - f_{AS} $. In the following, as we refer to the properties of the SW, $f_\mathrm{S}$ denotes the absolute value of the Stokes frequency, and wave vectors along that of the photons are assumed to be positive.

\begin{figure}
\includegraphics[width=8.5cm]{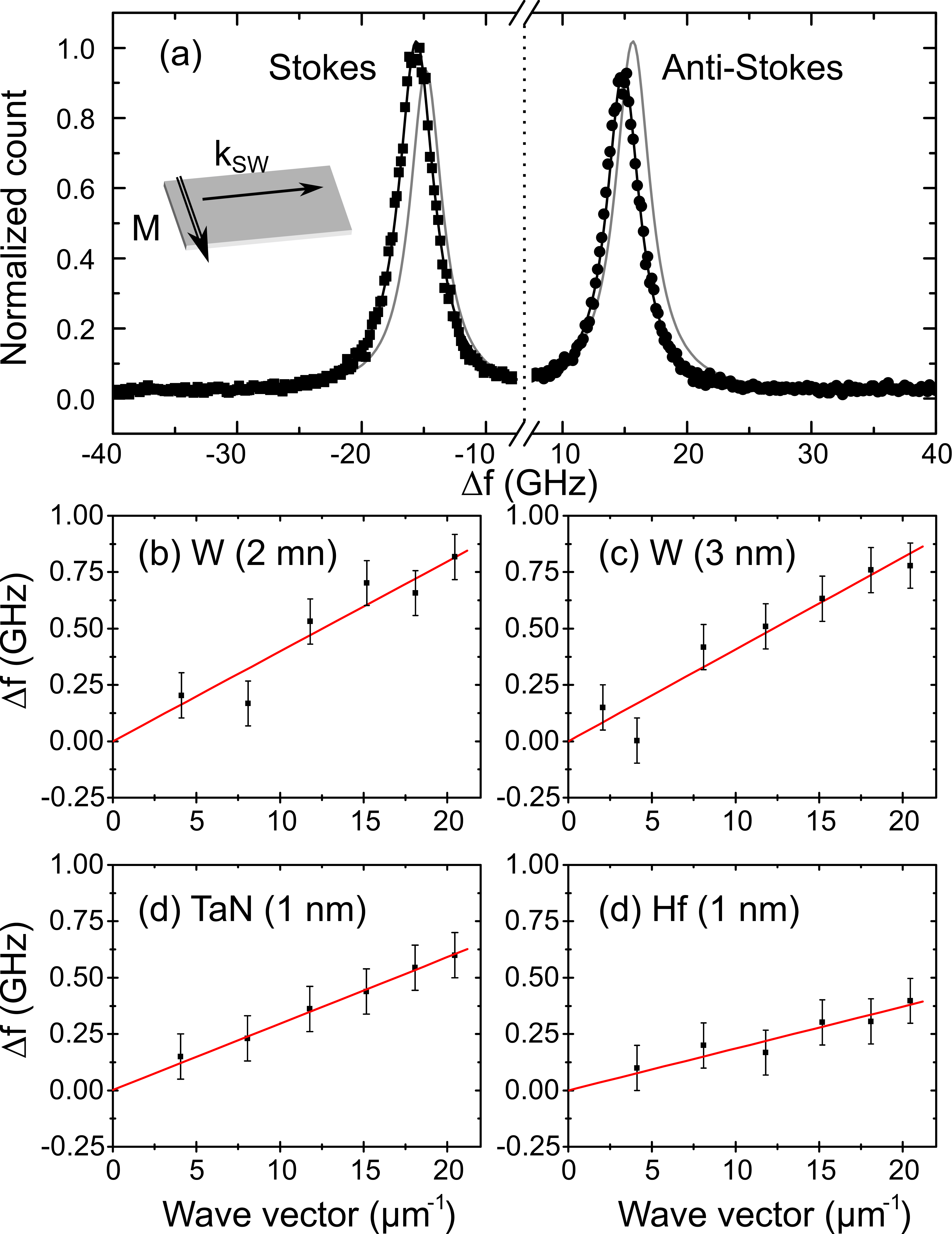}
\caption{(a) BLS spectra measured for the 2 nm thick W underlayer at a light incidence corresponding to $k_{\rm SW}$ = 20.45 $\mu$m$^{-1}$ and an in-plane field of 1T to saturate the magnetization. Lorentzian fit are also superposed to show the nonreciprocal behavior of the spin-wave, in gray the Stokes and anti-Stokes are inverted. The inset shows a schematic illustration of the Damon-Eshbach configuration used. (b)-(e) Linear fit of the frequency difference $\Delta f$ as a function of wave vector for the different underlayers. 
\label{Fig_BLS}}
\centering
\end{figure}

In Fig.~\ref{Fig_BLS}, we present results from our BLS measurements for the four different underlayers studied.
We observe a linear variation of the frequency nonreciprocity, $\Delta f$, with wave vector, for all the studied underlayers. The slopes obtained from a linear fit of this variation are summarized in Table~\ref{Tab_Hoffest}. The shifts are similar in magnitude for the two thicknesses of W and for TaN, while the nonreciprocity is approximately half as strong for the CoFeB film on Hf. We note that the nominal thickness for all the CoFeB layers studied is 1 nm, which allows us to exclude contributions from surface anisotropy~\cite{Stashkevich:2015} as the dominant mechanism for the nonreciprocity observed. Indeed, magnetostatic surface spin waves, which appear in thicker ferromagnetic films, can exhibit a frequency nonreciprocity even in the absence of a chiral interaction because they are localized to the film surfaces and probe different surfaces depending on their direction of propagation (e.g., left propagating waves are localized to the top surface while right propagation waves are localized to the bottom surface). In this way, a difference in surface anisotropies at the top and bottom film surfaces can induce a frequency nonreciprocity, but such effects should be negligible in the ultrathin films since $k_{\rm sw} t \ll 1$, where $t$ is the ferromagnetic film thickness. Moreover, the nonreciprocities observed are larger than what might be expected from a difference in surface anisotropies alone~\cite{Stashkevich:2015}.

%
%
\section{Discussion}

\begin{table*}
	\begin{ruledtabular}
	\begin{tabular}{cccccccccc}
 	Underlayer & $\mu_0 H_{\rm offset}$ (mT) & $D_{\rm creep}$ (mJ/m$^2$) & $\Delta f / k_{\rm SW}$ (MHz $\mu$m) & $D_{\rm BLS}$ (mJ/m$^2$) & $D_c$ (mJ/m$^2$) & $M_s$ (kA/m)~\cite{Natcomtorrejon2014} & $\alpha$\\
\hline
		W (2 nm)& $35 \pm 5$&$0.23 \pm 0.03$&$40 \pm 6$&$0.25 \pm 0.04$&$0.10$&$729$&$0.039 \pm 0.003$\\
		W (3 nm )& $15 \pm 5$&$0.12 \pm 0.03$&$41 \pm 6$&$0.27 \pm 0.04$&$0.11$&$788$&$0.033 \pm 0.006$\\
		Ta$_{48}$N$_{52}$ (1 nm)& $5 \pm 2$&$0.05 \pm 0.02$&$30 \pm 6$&$0.31 \pm 0.06$&$0.26$&$1235$&$0.015 \pm 0.003$\\
		Hf (1 nm) & $2 \pm 2$&$0.01 \pm 0.01$&$19 \pm 5$&$0.15 \pm 0.04$&$0.16$&$965$&$0.023 \pm 0.003$
 	\end{tabular}
 	\end{ruledtabular}
	\caption{\label{Tab_Hoffest} Measured in-plane field value leading to a minimum for the DW velocity for different underlayer composition and the slope of the non reciprocity frequency versus the wave vector measured by BLS. Extracted DMI obtained by the creep method is compared to BLS measurement and the critical DMI ($D_c$) leading to a full rotation of the DW core toward N\'eel configuration. The measured magnetization and the damping parameter are also given.}
\end{table*}

In this section, we discuss and contrast the estimates of the DMI obtained using the two experimental methods employed. We first begin by discussing results from the BLS experiment. As we mentioned above, the interfacial DMI results in a frequency nonreciprocity in PMA materials when the equilibrium magnetization is tilted away from the film normal~\cite{Moon:2013}. Advantage has already been taken of this behavior in different ultrathin film systems, where the amplitude and sign of the DMI constant, $D$ has been deduced from the measured frequency nonreciprocity~\cite{Di:2015, Belmeguenai:2015, Cho:2015}. This nonreciprocity is characterized by the difference in frequency between the Stokes and anti-Stokes peaks, which depends on $D$ through the relationship~\cite{Di:2015}
\begin{equation}
	\Delta f = \frac{2 \gamma}{\pi M_s} D k_{\rm sw},
\end{equation}
where $\gamma$ is the gyromagnetic constant and $M_{s}$ is the saturation magnetization. From the slope of the $\Delta f(k_{\rm SW})$ curve, we can therefore have a direct estimate of $D$ by assuming the bulk value of $\gamma$ and by using the values of $M_{s}$ obtained elsewhere~\cite{Natcomtorrejon2014}. The estimated values of $D$ from BLS measurements, $D_{\rm BLS}$, are reported in Table~\ref{Tab_Hoffest}. For all samples studied, we found that the Stokes frequencies are larger than the anti-Stokes frequencies, which indicates a positive DMI constant $D >0$ that favors a right-handed chirality~\cite{Belmeguenai:2015}.

Let us now turn our attention to our results of domain wall displacement in the creep regime. In PMA films, the wall motion at low fields in a disordered material is described by the creep model, in which the wall moves by series of thermally-activated correlated jump between successive pinning centers~\cite{Lemerle:1998, Chauve:2000}. The wall motion in this regime results from a competition between the disorder energy and the elastic energy associated with the domain wall, where the former favors a roughening of the wall by adapting to spatial profile of defects in the sample, while the latter favors a straight wall by minimizing the total length of the domain wall. In the present case, the wall energy is further modified by two additional competing energies: the Dzyaloshinskii-Moriya interaction, which favors a particular chirality of the domain wall, and the Zeeman energy associated with the in-plane applied-field, which can favor a different chirality depending on its orientation. To account for these additional terms, it has been proposed that the creep model can be modified by including a field-dependent wall energy $\sigma(H_{x})$~\cite{Je:2013}, 
\begin{equation}
	v_x(H_z, H_x) = v_0 \exp \left[ -\left( \frac{\alpha \left[ \sigma(H_x) \right]}{H_z} \right)^{1/4} \right].
\end{equation}
Here, the domain wall energy plays the role of an elastic energy and it can be estimated from the one-dimensional domain wall model~\cite{Lemerle:1998, Chauve:2000}, which can be modified to include the Zeeman and DMI terms~\cite{Je:2013}. This model predicts that any applied in-plane field will only modify the velocity through changes in the wall energy, $\sigma(H_{x})$. 
In the one-dimensional picture of domain wall motion, the DMI influences the domain wall motion as an effective magnetic field,
\begin{equation}
	H_{\rm DMI}=\frac{D}{\mu_0 M_s \lambda},
    \label{eq:DMI}
\end{equation}
where $\lambda$ is the domain wall width parameter $\sqrt{A/K_{\rm eff}}$, $A$ being the exchange constant, and $K_{\rm eff}$ the effective perpendicular anisotropy constant that takes into account the demagnetizing fields. This DMI field appears as an offset for the DW energy when we apply an in-plane field. With a non-zero DMI we expect a chiral magnetic N\'eel wall\cite{Thiaville:2012}. Since the domain wall velocity is directly linked to the domain wall energy $\sigma$ in the creep regime, one may expect that $H_{\rm DMI}$ also plays a role as an offset for the velocity curves shown in Fig.~\ref{Fig_Hip}.  We have verified using micromagnetic simulations~\cite{Vansteenkiste:2014} that the maximum in the domain wall energy is indeed obtained for the DMI effective field (at least within a one-dimensional approximation for the domain wall). We have also used these simulations to determine the critical value $D = D_c$ at which the equilibrium domain wall profile is a homochiral N\'eel wall, assuming an exchange stiffness of $A = 22.5$ pJ/m \cite{Burrowes:2013}. This transition is governed by the cost in dipolar energy that must be overcome by the DMI to transform a Bloch-type wall, which possess no volume magnetic charges, to a N\'eel type. The $D_c$ values are given in Table~\ref{Tab_Hoffest} for the different systems as a point of comparison.

By equating the offset fields, $\mu_0 H_{\rm offset}$ (Table~\ref{Tab_Hoffest}), measured from the global minimum of the the $v_x(H_x)$ curves, to the DMI effective field defined in Eq.~\ref{eq:DMI}, we obtain an estimate of the DMI constant $D$ from the domain wall measurements, which are reported in Table~\ref{Tab_Hoffest} as $D_{\rm creep}$. These values are consistent with the behavior of the $v_y(H_x)$ curves (black curves in Fig.~\ref{Fig_Hip}). Since DMI interaction tends to align the magnetization of the DW core in the N\'eel configuration, when if an in-plane field is applied parallel to a DW, favoring a Bloch wall, the magnetization in the DW core will converge faster to a Bloch configuration if the DMI is small. In terms of velocity, the stronger the DMI is, the softer the rounding of the V-shaped curve.

From Table~\ref{Tab_Hoffest}, we can draw the striking conclusion that the agreement between the values of $D$ obtained by domain wall creep and BLS depends strongly on the underlayer. BLS measurements indicate that for all the samples $D_{\rm BLS} \geq D_c > 0$, therefore right-handed N{\'e}el walls are expected for all the different underlayers studied. This correlates well with the two thicknesses of the W underlayer, for which both $v_x(H_x)$ and $v_y(H_x)$ curves can be explained with the simple creep model presented above. For the Hf and TaN underlayers, on the other hand, the agreement between the two methods is very poor, where the minima in the $v_x(H_x)$ curves suggest that domain walls in these samples should be closer to Bloch-type walls. It should be noted that the velocity curves for these two cases exhibit features that are not accounted for in the creep model, such as the local minimum located at $\mu_0 H_x = -55 \pm 10$~mT for the Hf underlayer.

One possible explanation of this discrepancy relates to the domain wall dynamics in the creep regime. This is neglected in usual treatments where focus is on the domain wall and pinning energies, which appear in the exponential factor in Eq.~\ref{eq:creep}. The dynamics is partly captured in the velocity prefactor, $v_0$, which has been shown to exhibit a nontrivial variation as a function of applied in-plane fields~\cite{Lavrijsen:2015}. Another related aspect of the dynamics involves the Gilbert damping of the domain wall. In Table~\ref{Tab_Hoffest}, we have included estimates of the Gilbert damping constant $\alpha$ that have been determined from vector-network-analyzer ferromagnetic resonance measurements on the films studied. Indeed, one can observe a clear correlation between the level of quantitative agreement between the BLS and domain wall creep measurements for $D$, which is very good for the largest value of $\alpha$ (W, 2 nm) and very poor for the smallest value of $\alpha$ (TaN). This trend strongly suggests that dynamical processes in weakly damped systems can give rise to behavior that is not captured by the simple creep model in which the in-plane applied field and DMI only act on the wall energy. A manifestation of such dynamical processes can be seen directly in Fig.~\ref{Fig_Kerr}, where the weakly damped systems (TaN and Hf) exhibit dendritic domain growth that is absent in the more strongly damped systems (W, 2 and 3 nm). We hypothesize that this is related to roughening of domain walls during motion in PMA films with low damping, which has been observed previously in different studies~\cite{Yamada:2011, Burrowes:2013}. We note that the effective damping parameter experienced by the domain wall may differ from the value obtained by ferromagnetic resonance~\cite{Weindler2014}, but the overall trend should persist since the differences due to nonlocal damping are inversely proportional to the domain wall width~\cite{Kim:2015}, which is similar for the different underlayers studied.

Another possible explanation is that Brillouin light scattering and measurements of domain wall creep probe two very different processes in magnetic systems. In BLS, thermally-populated long wavelength spin waves (with wavelengths larger than 270~nm) are probed over areas of 100~$\mu$m, which corresponds to the typical laser spot size used. As such, estimates of the DMI obtained reflect its strength averaged out over these distances. On the other hand, domain wall creep involves probing the competition between the (elastic) wall energy and local pinning potentials, which concern length scales of the order of the domain wall width~\cite{Lemerle:1998}, i.e., about 10-20 nm in our films. Moreover, creep motion necessarily involves thermally-activated jumps or avalanches between pinning sites, and it is possible that the DMI is systematically reduced at these sites by virtue of their existence. In other words, if we suppose that domain wall pinning systematically involves some combination of interface roughness, grain boundaries, and non-magnetic substitutional disorder in the ferromagnetic film --- regions in which the DMI is reduced because the spin-orbit--mediated coupling through the heavy metal underlayer is diminished --- then it follows that creep motion only involves local regions of weak DMI. This idea appears to be corroborated by recent experiments involving steady-state domain wall motion over larger distances (as attained in the flow regime of propagation), which provide estimates of the DMI strength that show better agreement with values obtained by BLS~\cite{Natcomtorrejon2014, Vanatka:2015}. It is also consistent with recent studies using scanning-NV magnetometry, which have revealed that spatial variations in the domain wall structure (i.e., the degree to which a wall is Bloch- or N\'eel-like) can occur~\cite{Gross_NV}. This is likely due to a local variation in magnetic parameters that can lead to the discrepancy discussed here.

Finally, the DMI for the W samples estimated using the BLS and creep measurements quantitatively agrees with the that obtained from current driven DW velocity measurements~\cite{Natcomtorrejon2014}. All measurements return the same sign of DMI for the TaN samples but not for the Hf samples.  It should be noted that the Hf thickness, and as a consequence, the structure of Hf\cite{LiuHayashiAPL2015}, is different from samples used in the current induced DW velocity measurements~\cite{Natcomtorrejon2014}. The Hf is thin and predominantly amorphous for the samples studied here whereas it is thicker and forms hcp structure for the samples used to evaluate DMI using the DW velocity measurements\cite{Natcomtorrejon2014}. As a consequence the structure of the Hf may play a role in defining the sign and amplitude of the DMI.
%
%
%
%
%
%
\section{Conclusion}
We have performed a detailed study, by boh DW dynamics under in-plane and out-of-plane magnetic field together with BLS measurement, of the influence of W, TaN or Hf on the induced interfacial DMI in CoFeB/MgO samples. BLS measurements indicate that for all the samples DMI is large enough to favor full N\'eel wall with a right handedness, in contrast to previous findings\cite{Natcomtorrejon2014}. These conclusions are not in quantitative agreement with those inferred from creep domain wall motion measurements. Qualitative agreement is obtained on the sign of the DMI constant, hence on the chirality of the magnetic textures in the samples. With regard to the amplitude of DMI, creep domain wall measurements show that the strength varies with the thickness of the underlayer while being larger than the critical value to have full N\'eel domain walls.  In the case of Hf, and TaN samples, with a lower damping than the W samples, the DMI value is much smaller than the values obtained by BLS while velocity dependence on the in-plane magnetic field exhibit asymmetries and local minima that calls for a more robust description of the in plane magnetic fields and DMI in the creep regime. Discrepancies between the two methods show also that taking into account spatial inhomogeneities of the DMI might be a key to explain the experimental observations quantitatively.

\begin{acknowledgments}
The authors wish to thank S. Eimer for the design of the in-plane electromagnet and F. Garcia-Sanchez, A. Thiaville, J.-P. Tetienne, T. Hingant, I. Gross, L. J. Martinez and V. Jacques for fruitful discussions and revision of the manuscript. This work was partially supported by the Agence Nationale de la Recherche (France) under Contract No. ANR-14-CE26-0012 (ULTRASKY). RS acknowledges additional support from the LabEx NanoSaclay.
\end{acknowledgments}


\begin{thebibliography}{35}%
\makeatletter
\providecommand \@ifxundefined [1]{%
 \@ifx{#1\undefined}
}%
\providecommand \@ifnum [1]{%
 \ifnum #1\expandafter \@firstoftwo
 \else \expandafter \@secondoftwo
 \fi
}%
\providecommand \@ifx [1]{%
 \ifx #1\expandafter \@firstoftwo
 \else \expandafter \@secondoftwo
 \fi
}%
\providecommand \natexlab [1]{#1}%
\providecommand \enquote  [1]{``#1''}%
\providecommand \bibnamefont  [1]{#1}%
\providecommand \bibfnamefont [1]{#1}%
\providecommand \citenamefont [1]{#1}%
\providecommand \href@noop [0]{\@secondoftwo}%
\providecommand \href [0]{\begingroup \@sanitize@url \@href}%
\providecommand \@href[1]{\@@startlink{#1}\@@href}%
\providecommand \@@href[1]{\endgroup#1\@@endlink}%
\providecommand \@sanitize@url [0]{\catcode `\\12\catcode `\$12\catcode
  `\&12\catcode `\#12\catcode `\^12\catcode `\_12\catcode `\%12\relax}%
\providecommand \@@startlink[1]{}%
\providecommand \@@endlink[0]{}%
\providecommand \url  [0]{\begingroup\@sanitize@url \@url }%
\providecommand \@url [1]{\endgroup\@href {#1}{\urlprefix }}%
\providecommand \urlprefix  [0]{URL }%
\providecommand \Eprint [0]{\href }%
\providecommand \doibase [0]{http://dx.doi.org/}%
\providecommand \selectlanguage [0]{\@gobble}%
\providecommand \bibinfo  [0]{\@secondoftwo}%
\providecommand \bibfield  [0]{\@secondoftwo}%
\providecommand \translation [1]{[#1]}%
\providecommand \BibitemOpen [0]{}%
\providecommand \bibitemStop [0]{}%
\providecommand \bibitemNoStop [0]{.\EOS\space}%
\providecommand \EOS [0]{\spacefactor3000\relax}%
\providecommand \BibitemShut  [1]{\csname bibitem#1\endcsname}%
\let\auto@bib@innerbib\@empty
\bibitem [{\citenamefont {Fert}\ and\ \citenamefont {Levy}(1980)}]{Fert:1980}%
  \BibitemOpen
  \bibfield  {author} {\bibinfo {author} {\bibfnamefont {A.}~\bibnamefont
  {Fert}}\ and\ \bibinfo {author} {\bibfnamefont {P.~M.}\ \bibnamefont
  {Levy}},\ }\href {\doibase 10.1103/PhysRevLett.44.1538} {\bibfield  {journal}
  {\bibinfo  {journal} {Phys. Rev. Lett.}\ }\textbf {\bibinfo {volume} {44}},\
  \bibinfo {pages} {1538} (\bibinfo {year} {1980})}\BibitemShut {NoStop}%
\bibitem [{\citenamefont {Fert}(1990)}]{fert_magnetic_1990}%
  \BibitemOpen
  \bibfield  {author} {\bibinfo {author} {\bibfnamefont {A.}~\bibnamefont
  {Fert}},\ }\href {\doibase 10.4028/www.scientific.net/MSF.59-60.439}
  {\bibfield  {journal} {\bibinfo  {journal} {Mat. Sci. Forum}\ }\textbf
  {\bibinfo {volume} {59-60}},\ \bibinfo {pages} {439} (\bibinfo {year}
  {1990})}\BibitemShut {NoStop}%
\bibitem [{\citenamefont {Bode}\ \emph {et~al.}(2007)\citenamefont {Bode},
  \citenamefont {Heide}, \citenamefont {Von~Bergmann}, \citenamefont
  {Ferriani}, \citenamefont {Heinze}, \citenamefont {Bihlmayer}, \citenamefont
  {Kubetzka}, \citenamefont {Pietzsch}, \citenamefont {Bl{\"u}gel},\ and\
  \citenamefont {Wiesendanger}}]{Bode:2007}%
  \BibitemOpen
  \bibfield  {author} {\bibinfo {author} {\bibfnamefont {M.}~\bibnamefont
  {Bode}}, \bibinfo {author} {\bibfnamefont {M.}~\bibnamefont {Heide}},
  \bibinfo {author} {\bibfnamefont {K.}~\bibnamefont {Von~Bergmann}}, \bibinfo
  {author} {\bibfnamefont {P.}~\bibnamefont {Ferriani}}, \bibinfo {author}
  {\bibfnamefont {S.}~\bibnamefont {Heinze}}, \bibinfo {author} {\bibfnamefont
  {G.}~\bibnamefont {Bihlmayer}}, \bibinfo {author} {\bibfnamefont
  {A.}~\bibnamefont {Kubetzka}}, \bibinfo {author} {\bibfnamefont
  {O.}~\bibnamefont {Pietzsch}}, \bibinfo {author} {\bibfnamefont
  {S.}~\bibnamefont {Bl{\"u}gel}}, \ and\ \bibinfo {author} {\bibfnamefont
  {R.}~\bibnamefont {Wiesendanger}},\ }\href@noop {} {\bibfield  {journal}
  {\bibinfo  {journal} {Nature}\ }\textbf {\bibinfo {volume} {447}},\ \bibinfo
  {pages} {190} (\bibinfo {year} {2007})}\BibitemShut {NoStop}%
\bibitem [{\citenamefont {Heinze}\ \emph {et~al.}(2011)\citenamefont {Heinze},
  \citenamefont {von Bergmann}, \citenamefont {Menzel}, \citenamefont {Brede},
  \citenamefont {Kubetzka}, \citenamefont {Wiesendanger}, \citenamefont
  {Bihlmayer},\ and\ \citenamefont {Bl{\"u}gel}}]{heinze2011spontaneous}%
  \BibitemOpen
  \bibfield  {author} {\bibinfo {author} {\bibfnamefont {S.}~\bibnamefont
  {Heinze}}, \bibinfo {author} {\bibfnamefont {K.}~\bibnamefont {von
  Bergmann}}, \bibinfo {author} {\bibfnamefont {M.}~\bibnamefont {Menzel}},
  \bibinfo {author} {\bibfnamefont {J.}~\bibnamefont {Brede}}, \bibinfo
  {author} {\bibfnamefont {A.}~\bibnamefont {Kubetzka}}, \bibinfo {author}
  {\bibfnamefont {R.}~\bibnamefont {Wiesendanger}}, \bibinfo {author}
  {\bibfnamefont {G.}~\bibnamefont {Bihlmayer}}, \ and\ \bibinfo {author}
  {\bibfnamefont {S.}~\bibnamefont {Bl{\"u}gel}},\ }\href@noop {} {\bibfield
  {journal} {\bibinfo  {journal} {Nat. Phys.}\ }\textbf {\bibinfo {volume}
  {7}},\ \bibinfo {pages} {713} (\bibinfo {year} {2011})}\BibitemShut {NoStop}%
\bibitem [{\citenamefont {Romming}\ \emph {et~al.}(2013)\citenamefont
  {Romming}, \citenamefont {Hanneken}, \citenamefont {Menzel}, \citenamefont
  {Bickel}, \citenamefont {Wolter}, \citenamefont {Von~Bergmann}, \citenamefont
  {Kubetzka},\ and\ \citenamefont {Wiesendanger}}]{Romming:2013}%
  \BibitemOpen
  \bibfield  {author} {\bibinfo {author} {\bibfnamefont {N.}~\bibnamefont
  {Romming}}, \bibinfo {author} {\bibfnamefont {C.}~\bibnamefont {Hanneken}},
  \bibinfo {author} {\bibfnamefont {M.}~\bibnamefont {Menzel}}, \bibinfo
  {author} {\bibfnamefont {J.~E.}\ \bibnamefont {Bickel}}, \bibinfo {author}
  {\bibfnamefont {B.}~\bibnamefont {Wolter}}, \bibinfo {author} {\bibfnamefont
  {K.}~\bibnamefont {Von~Bergmann}}, \bibinfo {author} {\bibfnamefont
  {A.}~\bibnamefont {Kubetzka}}, \ and\ \bibinfo {author} {\bibfnamefont
  {R.}~\bibnamefont {Wiesendanger}},\ }\href@noop {} {\bibfield  {journal}
  {\bibinfo  {journal} {Science}\ }\textbf {\bibinfo {volume} {341}},\ \bibinfo
  {pages} {636} (\bibinfo {year} {2013})}\BibitemShut {NoStop}%
\bibitem [{\citenamefont {Moreau-Luchaire}\ \emph {et~al.}(2016)\citenamefont
  {Moreau-Luchaire}, \citenamefont {Moutafis}, \citenamefont {Reyren},
  \citenamefont {Sampaio}, \citenamefont {Vaz}, \citenamefont {Van~Horne},
  \citenamefont {Bouzehouane}, \citenamefont {Garcia}, \citenamefont
  {Deranlot}, \citenamefont {Warnicke}, \citenamefont {Wohlh{\"u}ter},
  \citenamefont {George}, \citenamefont {Weigand}, \citenamefont {Raabe},
  \citenamefont {Cros},\ and\ \citenamefont
  {Fert}}]{Moreau_luchaireCNNano2016}%
  \BibitemOpen
  \bibfield  {author} {\bibinfo {author} {\bibfnamefont {C.}~\bibnamefont
  {Moreau-Luchaire}}, \bibinfo {author} {\bibfnamefont {C.}~\bibnamefont
  {Moutafis}}, \bibinfo {author} {\bibfnamefont {N.}~\bibnamefont {Reyren}},
  \bibinfo {author} {\bibfnamefont {J.}~\bibnamefont {Sampaio}}, \bibinfo
  {author} {\bibfnamefont {C.~A.~F.}\ \bibnamefont {Vaz}}, \bibinfo {author}
  {\bibfnamefont {N.}~\bibnamefont {Van~Horne}}, \bibinfo {author}
  {\bibfnamefont {K.}~\bibnamefont {Bouzehouane}}, \bibinfo {author}
  {\bibfnamefont {K.}~\bibnamefont {Garcia}}, \bibinfo {author} {\bibfnamefont
  {C.}~\bibnamefont {Deranlot}}, \bibinfo {author} {\bibfnamefont
  {P.}~\bibnamefont {Warnicke}}, \bibinfo {author} {\bibfnamefont
  {P.}~\bibnamefont {Wohlh{\"u}ter}}, \bibinfo {author} {\bibfnamefont {J.-M.}\
  \bibnamefont {George}}, \bibinfo {author} {\bibfnamefont {M.}~\bibnamefont
  {Weigand}}, \bibinfo {author} {\bibfnamefont {J.}~\bibnamefont {Raabe}},
  \bibinfo {author} {\bibfnamefont {V.}~\bibnamefont {Cros}}, \ and\ \bibinfo
  {author} {\bibfnamefont {A.}~\bibnamefont {Fert}},\ }\href
  {http://dx.doi.org/10.1038/nnano.2015.313} {\bibfield  {journal} {\bibinfo
  {journal} {Nat. Nanotechnol.}\ }\textbf {\bibinfo {volume} {Advance Online
  Publication}},\ \bibinfo {pages} {doi:10.1038/nnano.2015.313} (\bibinfo
  {year} {2016})}\BibitemShut {NoStop}%
\bibitem [{\citenamefont {Boulle}\ \emph {et~al.}(2016)\citenamefont {Boulle},
  \citenamefont {Vogel}, \citenamefont {Yang}, \citenamefont {Pizzini},
  \citenamefont {de~Souza~Chaves}, \citenamefont {Locatelli}, \citenamefont
  {Mente{\c s}}, \citenamefont {Sala}, \citenamefont {Buda-Prejbeanu},
  \citenamefont {Klein}, \citenamefont {Belmeguenai}, \citenamefont
  {Roussign{\'e}}, \citenamefont {Stashkevich}, \citenamefont {Ch{\'e}rif},
  \citenamefont {Aballe}, \citenamefont {Foerster}, \citenamefont {Chshiev},
  \citenamefont {Auffret}, \citenamefont {Miron},\ and\ \citenamefont
  {Gaudin}}]{Boulle:2016}%
  \BibitemOpen
  \bibfield  {author} {\bibinfo {author} {\bibfnamefont {O.}~\bibnamefont
  {Boulle}}, \bibinfo {author} {\bibfnamefont {J.}~\bibnamefont {Vogel}},
  \bibinfo {author} {\bibfnamefont {H.}~\bibnamefont {Yang}}, \bibinfo {author}
  {\bibfnamefont {S.}~\bibnamefont {Pizzini}}, \bibinfo {author} {\bibfnamefont
  {D.}~\bibnamefont {de~Souza~Chaves}}, \bibinfo {author} {\bibfnamefont
  {A.}~\bibnamefont {Locatelli}}, \bibinfo {author} {\bibfnamefont {T.~O.}\
  \bibnamefont {Mente{\c s}}}, \bibinfo {author} {\bibfnamefont
  {A.}~\bibnamefont {Sala}}, \bibinfo {author} {\bibfnamefont {L.~D.}\
  \bibnamefont {Buda-Prejbeanu}}, \bibinfo {author} {\bibfnamefont
  {O.}~\bibnamefont {Klein}}, \bibinfo {author} {\bibfnamefont
  {M.}~\bibnamefont {Belmeguenai}}, \bibinfo {author} {\bibfnamefont
  {Y.}~\bibnamefont {Roussign{\'e}}}, \bibinfo {author} {\bibfnamefont
  {A.}~\bibnamefont {Stashkevich}}, \bibinfo {author} {\bibfnamefont {S.~M.}\
  \bibnamefont {Ch{\'e}rif}}, \bibinfo {author} {\bibfnamefont
  {L.}~\bibnamefont {Aballe}}, \bibinfo {author} {\bibfnamefont
  {M.}~\bibnamefont {Foerster}}, \bibinfo {author} {\bibfnamefont
  {M.}~\bibnamefont {Chshiev}}, \bibinfo {author} {\bibfnamefont
  {S.}~\bibnamefont {Auffret}}, \bibinfo {author} {\bibfnamefont {I.~M.}\
  \bibnamefont {Miron}}, \ and\ \bibinfo {author} {\bibfnamefont
  {G.}~\bibnamefont {Gaudin}},\ }\href
  {http://dx.doi.org/10.1038/nnano.2015.315} {\bibfield  {journal} {\bibinfo
  {journal} {Nat. Nanotechnol.}\ }\textbf {\bibinfo {volume} {Advance Online
  Publication}},\ \bibinfo {pages} {doi:10.1038/nnano.2015.315} (\bibinfo
  {year} {2016})}\BibitemShut {NoStop}%
\bibitem [{\citenamefont {{Woo}}\ \emph {et~al.}(2016)\citenamefont {{Woo}},
  \citenamefont {{Litzius}}, \citenamefont {{Kr{\"u}ger}}, \citenamefont
  {{Im}}, \citenamefont {{Caretta}}, \citenamefont {{Richter}}, \citenamefont
  {{Mann}}, \citenamefont {{Krone}}, \citenamefont {{Reeve}}, \citenamefont
  {{Weigand}}, \citenamefont {{Agrawal}}, \citenamefont {{Fischer}},
  \citenamefont {{Kl{\"a}ui}},\ and\ \citenamefont {{Beach}}}]{Woo:2016}%
  \BibitemOpen
  \bibfield  {author} {\bibinfo {author} {\bibfnamefont {S.}~\bibnamefont
  {{Woo}}}, \bibinfo {author} {\bibfnamefont {K.}~\bibnamefont {{Litzius}}},
  \bibinfo {author} {\bibfnamefont {B.}~\bibnamefont {{Kr{\"u}ger}}}, \bibinfo
  {author} {\bibfnamefont {M.-Y.}\ \bibnamefont {{Im}}}, \bibinfo {author}
  {\bibfnamefont {L.}~\bibnamefont {{Caretta}}}, \bibinfo {author}
  {\bibfnamefont {K.}~\bibnamefont {{Richter}}}, \bibinfo {author}
  {\bibfnamefont {M.}~\bibnamefont {{Mann}}}, \bibinfo {author} {\bibfnamefont
  {A.}~\bibnamefont {{Krone}}}, \bibinfo {author} {\bibfnamefont
  {R.}~\bibnamefont {{Reeve}}}, \bibinfo {author} {\bibfnamefont
  {M.}~\bibnamefont {{Weigand}}}, \bibinfo {author} {\bibfnamefont
  {P.}~\bibnamefont {{Agrawal}}}, \bibinfo {author} {\bibfnamefont
  {P.}~\bibnamefont {{Fischer}}}, \bibinfo {author} {\bibfnamefont
  {M.}~\bibnamefont {{Kl{\"a}ui}}}, \ and\ \bibinfo {author} {\bibfnamefont
  {G.~S.~D.}\ \bibnamefont {{Beach}}},\ }\href
  {http://dx.doi.org/10.1038/nmat4593} {\bibfield  {journal} {\bibinfo
  {journal} {Nat. Mater.}\ }\textbf {\bibinfo {volume} {Advanced Online
  Publication}},\ \bibinfo {pages} {doi:10.1038/nmat4593} (\bibinfo {year}
  {2016})}\BibitemShut {NoStop}%
\bibitem [{\citenamefont {Tetienne}\ \emph {et~al.}(2015)\citenamefont
  {Tetienne}, \citenamefont {Hingant}, \citenamefont {MartÃ­nez},
  \citenamefont {Rohart}, \citenamefont {Thiaville}, \citenamefont {Diez},
  \citenamefont {Garcia}, \citenamefont {Adam}, \citenamefont {Kim},
  \citenamefont {Roch}, \citenamefont {Miron}, \citenamefont {Gaudin},
  \citenamefont {Vila}, \citenamefont {Ocker}, \citenamefont {Ravelosona},\
  and\ \citenamefont {Jacques}}]{Tetienne2015}%
  \BibitemOpen
  \bibfield  {author} {\bibinfo {author} {\bibfnamefont {J.-P.}\ \bibnamefont
  {Tetienne}}, \bibinfo {author} {\bibfnamefont {T.}~\bibnamefont {Hingant}},
  \bibinfo {author} {\bibfnamefont {L.}~\bibnamefont {MartÃ­nez}}, \bibinfo
  {author} {\bibfnamefont {S.}~\bibnamefont {Rohart}}, \bibinfo {author}
  {\bibfnamefont {A.}~\bibnamefont {Thiaville}}, \bibinfo {author}
  {\bibfnamefont {L.~H.}\ \bibnamefont {Diez}}, \bibinfo {author}
  {\bibfnamefont {K.}~\bibnamefont {Garcia}}, \bibinfo {author} {\bibfnamefont
  {J.-P.}\ \bibnamefont {Adam}}, \bibinfo {author} {\bibfnamefont {J.-V.}\
  \bibnamefont {Kim}}, \bibinfo {author} {\bibfnamefont {J.-F.}\ \bibnamefont
  {Roch}}, \bibinfo {author} {\bibfnamefont {I.}~\bibnamefont {Miron}},
  \bibinfo {author} {\bibfnamefont {G.}~\bibnamefont {Gaudin}}, \bibinfo
  {author} {\bibfnamefont {L.}~\bibnamefont {Vila}}, \bibinfo {author}
  {\bibfnamefont {B.}~\bibnamefont {Ocker}}, \bibinfo {author} {\bibfnamefont
  {D.}~\bibnamefont {Ravelosona}}, \ and\ \bibinfo {author} {\bibfnamefont
  {V.}~\bibnamefont {Jacques}},\ }\href {http://dx.doi.org/10.1038/ncomms7733}
  {\bibfield  {journal} {\bibinfo  {journal} {Nat Commun}\ }\textbf {\bibinfo
  {volume} {6}},\  (\bibinfo {year} {2015})}\BibitemShut {NoStop}%
\bibitem [{\citenamefont {Benitez}\ \emph {et~al.}(2015)\citenamefont
  {Benitez}, \citenamefont {Hrabec}, \citenamefont {Mihai}, \citenamefont
  {Moore}, \citenamefont {Burnell}, \citenamefont {McGrouther}, \citenamefont
  {Marrows},\ and\ \citenamefont {McVitie}}]{Benitez:2015}%
  \BibitemOpen
  \bibfield  {author} {\bibinfo {author} {\bibfnamefont {M.~J.}\ \bibnamefont
  {Benitez}}, \bibinfo {author} {\bibfnamefont {A.}~\bibnamefont {Hrabec}},
  \bibinfo {author} {\bibfnamefont {A.~P.}\ \bibnamefont {Mihai}}, \bibinfo
  {author} {\bibfnamefont {T.~A.}\ \bibnamefont {Moore}}, \bibinfo {author}
  {\bibfnamefont {G.}~\bibnamefont {Burnell}}, \bibinfo {author} {\bibfnamefont
  {D.}~\bibnamefont {McGrouther}}, \bibinfo {author} {\bibfnamefont {C.~H.}\
  \bibnamefont {Marrows}}, \ and\ \bibinfo {author} {\bibfnamefont
  {S.}~\bibnamefont {McVitie}},\ }\href@noop {} {\bibfield  {journal} {\bibinfo
   {journal} {Nat. Commun.}\ }\textbf {\bibinfo {volume} {6}} (\bibinfo {year}
  {2015})}\BibitemShut {NoStop}%
\bibitem [{\citenamefont {Belmeguenai}\ \emph {et~al.}(2015)\citenamefont
  {Belmeguenai}, \citenamefont {Adam}, \citenamefont {Roussign\'e},
  \citenamefont {Eimer}, \citenamefont {Devolder}, \citenamefont {Kim},
  \citenamefont {Cherif}, \citenamefont {Stashkevich},\ and\ \citenamefont
  {Thiaville}}]{Belmeguenai:2015}%
  \BibitemOpen
  \bibfield  {author} {\bibinfo {author} {\bibfnamefont {M.}~\bibnamefont
  {Belmeguenai}}, \bibinfo {author} {\bibfnamefont {J.-P.}\ \bibnamefont
  {Adam}}, \bibinfo {author} {\bibfnamefont {Y.}~\bibnamefont {Roussign\'e}},
  \bibinfo {author} {\bibfnamefont {S.}~\bibnamefont {Eimer}}, \bibinfo
  {author} {\bibfnamefont {T.}~\bibnamefont {Devolder}}, \bibinfo {author}
  {\bibfnamefont {J.-V.}\ \bibnamefont {Kim}}, \bibinfo {author} {\bibfnamefont
  {S.~M.}\ \bibnamefont {Cherif}}, \bibinfo {author} {\bibfnamefont
  {A.}~\bibnamefont {Stashkevich}}, \ and\ \bibinfo {author} {\bibfnamefont
  {A.}~\bibnamefont {Thiaville}},\ }\href {\doibase 10.1103/PhysRevB.91.180405}
  {\bibfield  {journal} {\bibinfo  {journal} {Phys. Rev. B}\ }\textbf {\bibinfo
  {volume} {91}},\ \bibinfo {pages} {180405} (\bibinfo {year}
  {2015})}\BibitemShut {NoStop}%
\bibitem [{\citenamefont {Freimuth}\ \emph {et~al.}(2014)\citenamefont
  {Freimuth}, \citenamefont {Bl{\"u}gel},\ and\ \citenamefont
  {Mokrousov}}]{Freimuth:2014}%
  \BibitemOpen
  \bibfield  {author} {\bibinfo {author} {\bibfnamefont {F.}~\bibnamefont
  {Freimuth}}, \bibinfo {author} {\bibfnamefont {S.}~\bibnamefont
  {Bl{\"u}gel}}, \ and\ \bibinfo {author} {\bibfnamefont {Y.}~\bibnamefont
  {Mokrousov}},\ }\href@noop {} {\bibfield  {journal} {\bibinfo  {journal} {J.
  Phys. Condens. Matter}\ }\textbf {\bibinfo {volume} {26}},\ \bibinfo {pages}
  {104202} (\bibinfo {year} {2014})}\BibitemShut {NoStop}%
\bibitem [{\citenamefont {Jiang}\ \emph {et~al.}(2015)\citenamefont {Jiang},
  \citenamefont {Upadhyaya}, \citenamefont {Zhang}, \citenamefont {Yu},
  \citenamefont {Jungfleisch}, \citenamefont {Fradin}, \citenamefont {Pearson},
  \citenamefont {Tserkovnyak}, \citenamefont {Wang}, \citenamefont {Heinonen},
  \citenamefont {te~Velthuis},\ and\ \citenamefont {Hoffmann}}]{Jiang:2015}%
  \BibitemOpen
  \bibfield  {author} {\bibinfo {author} {\bibfnamefont {W.}~\bibnamefont
  {Jiang}}, \bibinfo {author} {\bibfnamefont {P.}~\bibnamefont {Upadhyaya}},
  \bibinfo {author} {\bibfnamefont {W.}~\bibnamefont {Zhang}}, \bibinfo
  {author} {\bibfnamefont {G.}~\bibnamefont {Yu}}, \bibinfo {author}
  {\bibfnamefont {M.~B.}\ \bibnamefont {Jungfleisch}}, \bibinfo {author}
  {\bibfnamefont {F.~Y.}\ \bibnamefont {Fradin}}, \bibinfo {author}
  {\bibfnamefont {J.~E.}\ \bibnamefont {Pearson}}, \bibinfo {author}
  {\bibfnamefont {Y.}~\bibnamefont {Tserkovnyak}}, \bibinfo {author}
  {\bibfnamefont {K.~L.}\ \bibnamefont {Wang}}, \bibinfo {author}
  {\bibfnamefont {O.}~\bibnamefont {Heinonen}}, \bibinfo {author}
  {\bibfnamefont {S.~G.~E.}\ \bibnamefont {te~Velthuis}}, \ and\ \bibinfo
  {author} {\bibfnamefont {A.}~\bibnamefont {Hoffmann}},\ }\href@noop {}
  {\bibfield  {journal} {\bibinfo  {journal} {Science}\ }\textbf {\bibinfo
  {volume} {349}},\ \bibinfo {pages} {283} (\bibinfo {year}
  {2015})}\BibitemShut {NoStop}%
\bibitem [{\citenamefont {Torrejon}\ \emph {et~al.}(2014)\citenamefont
  {Torrejon}, \citenamefont {Kim}, \citenamefont {Sinha}, \citenamefont
  {Mitani}, \citenamefont {Hayashi}, \citenamefont {Yamanouchi},\ and\
  \citenamefont {Ohno}}]{Natcomtorrejon2014}%
  \BibitemOpen
  \bibfield  {author} {\bibinfo {author} {\bibfnamefont {J.}~\bibnamefont
  {Torrejon}}, \bibinfo {author} {\bibfnamefont {J.}~\bibnamefont {Kim}},
  \bibinfo {author} {\bibfnamefont {J.}~\bibnamefont {Sinha}}, \bibinfo
  {author} {\bibfnamefont {S.}~\bibnamefont {Mitani}}, \bibinfo {author}
  {\bibfnamefont {M.}~\bibnamefont {Hayashi}}, \bibinfo {author} {\bibfnamefont
  {M.}~\bibnamefont {Yamanouchi}}, \ and\ \bibinfo {author} {\bibfnamefont
  {H.}~\bibnamefont {Ohno}},\ }\href {http://dx.doi.org/10.1038/ncomms5655}
  {\bibfield  {journal} {\bibinfo  {journal} {Nat. Commun.}\ }\textbf {\bibinfo
  {volume} {5}},\ \bibinfo {pages} {4655} (\bibinfo {year} {2014})}\BibitemShut
  {NoStop}%
\bibitem [{\citenamefont {Je}\ \emph {et~al.}(2013)\citenamefont {Je},
  \citenamefont {Kim}, \citenamefont {Yoo}, \citenamefont {Min}, \citenamefont
  {Lee},\ and\ \citenamefont {Choe}}]{Je:2013}%
  \BibitemOpen
  \bibfield  {author} {\bibinfo {author} {\bibfnamefont {S.-G.}\ \bibnamefont
  {Je}}, \bibinfo {author} {\bibfnamefont {D.-H.}\ \bibnamefont {Kim}},
  \bibinfo {author} {\bibfnamefont {S.-C.}\ \bibnamefont {Yoo}}, \bibinfo
  {author} {\bibfnamefont {B.-C.}\ \bibnamefont {Min}}, \bibinfo {author}
  {\bibfnamefont {K.-J.}\ \bibnamefont {Lee}}, \ and\ \bibinfo {author}
  {\bibfnamefont {S.-B.}\ \bibnamefont {Choe}},\ }\href {\doibase
  10.1103/PhysRevB.88.214401} {\bibfield  {journal} {\bibinfo  {journal} {Phys.
  Rev. B}\ }\textbf {\bibinfo {volume} {88}},\ \bibinfo {pages} {214401}
  (\bibinfo {year} {2013})}\BibitemShut {NoStop}%
\bibitem [{\citenamefont {Hrabec}\ \emph {et~al.}(2014)\citenamefont {Hrabec},
  \citenamefont {Porter}, \citenamefont {Wells}, \citenamefont {Benitez},
  \citenamefont {Burnell}, \citenamefont {McVitie}, \citenamefont {McGrouther},
  \citenamefont {Moore},\ and\ \citenamefont {Marrows}}]{Hrabec:2014}%
  \BibitemOpen
  \bibfield  {author} {\bibinfo {author} {\bibfnamefont {A.}~\bibnamefont
  {Hrabec}}, \bibinfo {author} {\bibfnamefont {N.~A.}\ \bibnamefont {Porter}},
  \bibinfo {author} {\bibfnamefont {A.}~\bibnamefont {Wells}}, \bibinfo
  {author} {\bibfnamefont {M.~J.}\ \bibnamefont {Benitez}}, \bibinfo {author}
  {\bibfnamefont {G.}~\bibnamefont {Burnell}}, \bibinfo {author} {\bibfnamefont
  {S.}~\bibnamefont {McVitie}}, \bibinfo {author} {\bibfnamefont
  {D.}~\bibnamefont {McGrouther}}, \bibinfo {author} {\bibfnamefont {T.~A.}\
  \bibnamefont {Moore}}, \ and\ \bibinfo {author} {\bibfnamefont {C.~H.}\
  \bibnamefont {Marrows}},\ }\href {\doibase 10.1103/PhysRevB.90.020402}
  {\bibfield  {journal} {\bibinfo  {journal} {Phys. Rev. B}\ }\textbf {\bibinfo
  {volume} {90}},\ \bibinfo {pages} {020402} (\bibinfo {year}
  {2014})}\BibitemShut {NoStop}%
\bibitem [{\citenamefont {Moon}\ \emph {et~al.}(2013)\citenamefont {Moon},
  \citenamefont {Seo}, \citenamefont {Lee}, \citenamefont {Kim}, \citenamefont
  {Ryu}, \citenamefont {Lee}, \citenamefont {McMichael},\ and\ \citenamefont
  {Stiles}}]{Moon:2013}%
  \BibitemOpen
  \bibfield  {author} {\bibinfo {author} {\bibfnamefont {J.-H.}\ \bibnamefont
  {Moon}}, \bibinfo {author} {\bibfnamefont {S.-M.}\ \bibnamefont {Seo}},
  \bibinfo {author} {\bibfnamefont {K.-J.}\ \bibnamefont {Lee}}, \bibinfo
  {author} {\bibfnamefont {K.-W.}\ \bibnamefont {Kim}}, \bibinfo {author}
  {\bibfnamefont {J.}~\bibnamefont {Ryu}}, \bibinfo {author} {\bibfnamefont
  {H.-W.}\ \bibnamefont {Lee}}, \bibinfo {author} {\bibfnamefont {R.~D.}\
  \bibnamefont {McMichael}}, \ and\ \bibinfo {author} {\bibfnamefont {M.~D.}\
  \bibnamefont {Stiles}},\ }\href
  {http://link.aps.org/doi/10.1103/PhysRevB.88.184404} {\bibfield  {journal}
  {\bibinfo  {journal} {Phys. Rev. B}\ }\textbf {\bibinfo {volume} {88}},\
  \bibinfo {pages} {184404} (\bibinfo {year} {2013})}\BibitemShut {NoStop}%
\bibitem [{\citenamefont {Di}\ \emph {et~al.}(2015)\citenamefont {Di},
  \citenamefont {Zhang}, \citenamefont {Lim}, \citenamefont {Ng}, \citenamefont
  {Kuok}, \citenamefont {Yu}, \citenamefont {Yoon}, \citenamefont {Qiu},\ and\
  \citenamefont {Yang}}]{Di:2015}%
  \BibitemOpen
  \bibfield  {author} {\bibinfo {author} {\bibfnamefont {K.}~\bibnamefont
  {Di}}, \bibinfo {author} {\bibfnamefont {V.~L.}\ \bibnamefont {Zhang}},
  \bibinfo {author} {\bibfnamefont {H.~S.}\ \bibnamefont {Lim}}, \bibinfo
  {author} {\bibfnamefont {S.~C.}\ \bibnamefont {Ng}}, \bibinfo {author}
  {\bibfnamefont {M.~H.}\ \bibnamefont {Kuok}}, \bibinfo {author}
  {\bibfnamefont {J.}~\bibnamefont {Yu}}, \bibinfo {author} {\bibfnamefont
  {J.}~\bibnamefont {Yoon}}, \bibinfo {author} {\bibfnamefont {X.}~\bibnamefont
  {Qiu}}, \ and\ \bibinfo {author} {\bibfnamefont {H.}~\bibnamefont {Yang}},\
  }\href {\doibase 10.1103/PhysRevLett.114.047201} {\bibfield  {journal}
  {\bibinfo  {journal} {Phys. Rev. Lett.}\ }\textbf {\bibinfo {volume} {114}},\
  \bibinfo {pages} {047201} (\bibinfo {year} {2015})}\BibitemShut {NoStop}%
\bibitem [{\citenamefont {Liu}\ \emph {et~al.}(2015)\citenamefont {Liu},
  \citenamefont {Ohkubo}, \citenamefont {Mitani}, \citenamefont {Hono},\ and\
  \citenamefont {Hayashi}}]{LiuHayashiAPL2015}%
  \BibitemOpen
  \bibfield  {author} {\bibinfo {author} {\bibfnamefont {J.}~\bibnamefont
  {Liu}}, \bibinfo {author} {\bibfnamefont {T.}~\bibnamefont {Ohkubo}},
  \bibinfo {author} {\bibfnamefont {S.}~\bibnamefont {Mitani}}, \bibinfo
  {author} {\bibfnamefont {K.}~\bibnamefont {Hono}}, \ and\ \bibinfo {author}
  {\bibfnamefont {M.}~\bibnamefont {Hayashi}},\ }\href {\doibase
  http://dx.doi.org/10.1063/1.4937452} {\bibfield  {journal} {\bibinfo
  {journal} {Applied Physics Letters}\ }\textbf {\bibinfo {volume} {107}},\
  \bibinfo {eid} {232408} (\bibinfo {year} {2015}),\
  http://dx.doi.org/10.1063/1.4937452}\BibitemShut {NoStop}%
\bibitem [{\citenamefont {Metaxas}\ \emph {et~al.}(2007)\citenamefont
  {Metaxas}, \citenamefont {Jamet}, \citenamefont {Mougin}, \citenamefont
  {Cormier}, \citenamefont {Ferr\'e}, \citenamefont {Baltz}, \citenamefont
  {Rodmacq}, \citenamefont {Dieny},\ and\ \citenamefont
  {Stamps}}]{Metaxas2007}%
  \BibitemOpen
  \bibfield  {author} {\bibinfo {author} {\bibfnamefont {P.~J.}\ \bibnamefont
  {Metaxas}}, \bibinfo {author} {\bibfnamefont {J.~P.}\ \bibnamefont {Jamet}},
  \bibinfo {author} {\bibfnamefont {A.}~\bibnamefont {Mougin}}, \bibinfo
  {author} {\bibfnamefont {M.}~\bibnamefont {Cormier}}, \bibinfo {author}
  {\bibfnamefont {J.}~\bibnamefont {Ferr\'e}}, \bibinfo {author} {\bibfnamefont
  {V.}~\bibnamefont {Baltz}}, \bibinfo {author} {\bibfnamefont
  {B.}~\bibnamefont {Rodmacq}}, \bibinfo {author} {\bibfnamefont
  {B.}~\bibnamefont {Dieny}}, \ and\ \bibinfo {author} {\bibfnamefont {R.~L.}\
  \bibnamefont {Stamps}},\ }\href {\doibase 10.1103/PhysRevLett.99.217208}
  {\bibfield  {journal} {\bibinfo  {journal} {Phys. Rev. Lett.}\ }\textbf
  {\bibinfo {volume} {99}},\ \bibinfo {pages} {217208} (\bibinfo {year}
  {2007})}\BibitemShut {NoStop}%
\bibitem [{\citenamefont {Lemerle}\ \emph {et~al.}(1998)\citenamefont
  {Lemerle}, \citenamefont {Ferr\'e}, \citenamefont {Chappert}, \citenamefont
  {Mathet}, \citenamefont {Giamarchi},\ and\ \citenamefont
  {Le~Doussal}}]{Lemerle:1998}%
  \BibitemOpen
  \bibfield  {author} {\bibinfo {author} {\bibfnamefont {S.}~\bibnamefont
  {Lemerle}}, \bibinfo {author} {\bibfnamefont {J.}~\bibnamefont {Ferr\'e}},
  \bibinfo {author} {\bibfnamefont {C.}~\bibnamefont {Chappert}}, \bibinfo
  {author} {\bibfnamefont {V.}~\bibnamefont {Mathet}}, \bibinfo {author}
  {\bibfnamefont {T.}~\bibnamefont {Giamarchi}}, \ and\ \bibinfo {author}
  {\bibfnamefont {P.}~\bibnamefont {Le~Doussal}},\ }\href {\doibase
  10.1103/PhysRevLett.80.849} {\bibfield  {journal} {\bibinfo  {journal} {Phys.
  Rev. Lett.}\ }\textbf {\bibinfo {volume} {80}},\ \bibinfo {pages} {849}
  (\bibinfo {year} {1998})}\BibitemShut {NoStop}%
\bibitem [{\citenamefont {Kabanov}\ \emph {et~al.}(2010)\citenamefont
  {Kabanov}, \citenamefont {Iunin}, \citenamefont {Nikitenko}, \citenamefont
  {Shapiro}, \citenamefont {Shull}, \citenamefont {Zhu},\ and\ \citenamefont
  {Chien}}]{Kabanov:2010}%
  \BibitemOpen
  \bibfield  {author} {\bibinfo {author} {\bibfnamefont {Y.~P.}\ \bibnamefont
  {Kabanov}}, \bibinfo {author} {\bibfnamefont {Y.~L.}\ \bibnamefont {Iunin}},
  \bibinfo {author} {\bibfnamefont {V.~I.}\ \bibnamefont {Nikitenko}}, \bibinfo
  {author} {\bibfnamefont {A.~J.}\ \bibnamefont {Shapiro}}, \bibinfo {author}
  {\bibfnamefont {R.~D.}\ \bibnamefont {Shull}}, \bibinfo {author}
  {\bibfnamefont {L.~Y.}\ \bibnamefont {Zhu}}, \ and\ \bibinfo {author}
  {\bibfnamefont {C.~L.}\ \bibnamefont {Chien}},\ }\href@noop {} {\bibfield
  {journal} {\bibinfo  {journal} {IEEE Trans. Magn.}\ }\textbf {\bibinfo
  {volume} {46}},\ \bibinfo {pages} {2220} (\bibinfo {year}
  {2010})}\BibitemShut {NoStop}%
\bibitem [{\citenamefont {Lavrijsen}\ \emph {et~al.}(2015)\citenamefont
  {Lavrijsen}, \citenamefont {Hartmann}, \citenamefont {van~den Brink},
  \citenamefont {Yin}, \citenamefont {Barcones}, \citenamefont {Duine},
  \citenamefont {Verheijen}, \citenamefont {Swagten},\ and\ \citenamefont
  {Koopmans}}]{Lavrijsen:2015}%
  \BibitemOpen
  \bibfield  {author} {\bibinfo {author} {\bibfnamefont {R.}~\bibnamefont
  {Lavrijsen}}, \bibinfo {author} {\bibfnamefont {D.~M.~F.}\ \bibnamefont
  {Hartmann}}, \bibinfo {author} {\bibfnamefont {A.}~\bibnamefont {van~den
  Brink}}, \bibinfo {author} {\bibfnamefont {Y.}~\bibnamefont {Yin}}, \bibinfo
  {author} {\bibfnamefont {B.}~\bibnamefont {Barcones}}, \bibinfo {author}
  {\bibfnamefont {R.~A.}\ \bibnamefont {Duine}}, \bibinfo {author}
  {\bibfnamefont {M.~A.}\ \bibnamefont {Verheijen}}, \bibinfo {author}
  {\bibfnamefont {H.~J.~M.}\ \bibnamefont {Swagten}}, \ and\ \bibinfo {author}
  {\bibfnamefont {B.}~\bibnamefont {Koopmans}},\ }\href {\doibase
  10.1103/PhysRevB.91.104414} {\bibfield  {journal} {\bibinfo  {journal} {Phys.
  Rev. B}\ }\textbf {\bibinfo {volume} {91}},\ \bibinfo {pages} {104414}
  (\bibinfo {year} {2015})}\BibitemShut {NoStop}%
\bibitem [{\citenamefont {Va\v{n}atka}\ \emph {et~al.}(2015)\citenamefont
  {Va\v{n}atka}, \citenamefont {Rojas~S{\'a}nchez}, \citenamefont {Vogel},
  \citenamefont {Bonfim}, \citenamefont {Belmeguenai}, \citenamefont
  {Roussign{\'e}}, \citenamefont {Stashkevich}, \citenamefont {Thiaville},\
  and\ \citenamefont {Pizzini}}]{Vanatka:2015}%
  \BibitemOpen
  \bibfield  {author} {\bibinfo {author} {\bibfnamefont {M.}~\bibnamefont
  {Va\v{n}atka}}, \bibinfo {author} {\bibfnamefont {J.~C.}\ \bibnamefont
  {Rojas~S{\'a}nchez}}, \bibinfo {author} {\bibfnamefont {J.}~\bibnamefont
  {Vogel}}, \bibinfo {author} {\bibfnamefont {M.}~\bibnamefont {Bonfim}},
  \bibinfo {author} {\bibfnamefont {M.}~\bibnamefont {Belmeguenai}}, \bibinfo
  {author} {\bibfnamefont {Y.}~\bibnamefont {Roussign{\'e}}}, \bibinfo {author}
  {\bibfnamefont {A.}~\bibnamefont {Stashkevich}}, \bibinfo {author}
  {\bibfnamefont {A.}~\bibnamefont {Thiaville}}, \ and\ \bibinfo {author}
  {\bibfnamefont {S.}~\bibnamefont {Pizzini}},\ }\href@noop {} {\bibfield
  {journal} {\bibinfo  {journal} {J. Phys.: Condens. Matter}\ ,\ \bibinfo
  {pages} {326002}} (\bibinfo {year} {2015})}\BibitemShut {NoStop}%
\bibitem [{\citenamefont {Garcia-Sanchez}\ \emph {et~al.}(2014)\citenamefont
  {Garcia-Sanchez}, \citenamefont {Borys}, \citenamefont {Vansteenkiste},
  \citenamefont {Kim},\ and\ \citenamefont {Stamps}}]{Garcia-Sanchez:2014}%
  \BibitemOpen
  \bibfield  {author} {\bibinfo {author} {\bibfnamefont {F.}~\bibnamefont
  {Garcia-Sanchez}}, \bibinfo {author} {\bibfnamefont {P.}~\bibnamefont
  {Borys}}, \bibinfo {author} {\bibfnamefont {A.}~\bibnamefont
  {Vansteenkiste}}, \bibinfo {author} {\bibfnamefont {J.-V.}\ \bibnamefont
  {Kim}}, \ and\ \bibinfo {author} {\bibfnamefont {R.~L.}\ \bibnamefont
  {Stamps}},\ }\href {\doibase 10.1103/PhysRevB.89.224408} {\bibfield
  {journal} {\bibinfo  {journal} {Phys. Rev. B}\ }\textbf {\bibinfo {volume}
  {89}},\ \bibinfo {pages} {224408} (\bibinfo {year} {2014})}\BibitemShut
  {NoStop}%
\bibitem [{\citenamefont {Stashkevich}\ \emph {et~al.}(2015)\citenamefont
  {Stashkevich}, \citenamefont {Belmeguenai}, \citenamefont {Roussign\'e},
  \citenamefont {Cherif}, \citenamefont {Kostylev}, \citenamefont {Gabor},
  \citenamefont {Lacour}, \citenamefont {Tiusan},\ and\ \citenamefont
  {Hehn}}]{Stashkevich:2015}%
  \BibitemOpen
  \bibfield  {author} {\bibinfo {author} {\bibfnamefont {A.~A.}\ \bibnamefont
  {Stashkevich}}, \bibinfo {author} {\bibfnamefont {M.}~\bibnamefont
  {Belmeguenai}}, \bibinfo {author} {\bibfnamefont {Y.}~\bibnamefont
  {Roussign\'e}}, \bibinfo {author} {\bibfnamefont {S.~M.}\ \bibnamefont
  {Cherif}}, \bibinfo {author} {\bibfnamefont {M.}~\bibnamefont {Kostylev}},
  \bibinfo {author} {\bibfnamefont {M.}~\bibnamefont {Gabor}}, \bibinfo
  {author} {\bibfnamefont {D.}~\bibnamefont {Lacour}}, \bibinfo {author}
  {\bibfnamefont {C.}~\bibnamefont {Tiusan}}, \ and\ \bibinfo {author}
  {\bibfnamefont {M.}~\bibnamefont {Hehn}},\ }\href {\doibase
  10.1103/PhysRevB.91.214409} {\bibfield  {journal} {\bibinfo  {journal} {Phys.
  Rev. B}\ }\textbf {\bibinfo {volume} {91}},\ \bibinfo {pages} {214409}
  (\bibinfo {year} {2015})}\BibitemShut {NoStop}%
\bibitem [{\citenamefont {Cho}\ \emph {et~al.}(2015)\citenamefont {Cho},
  \citenamefont {Kim}, \citenamefont {Lee}, \citenamefont {Lavrijsen},
  \citenamefont {Solignac}, \citenamefont {Yin}, \citenamefont {Han},
  \citenamefont {van Hoof}, \citenamefont {Swagten}, \citenamefont {Koopmans},
  \citenamefont {Kim},\ and\ \citenamefont {You}}]{Cho:2015}%
  \BibitemOpen
  \bibfield  {author} {\bibinfo {author} {\bibfnamefont {J.}~\bibnamefont
  {Cho}}, \bibinfo {author} {\bibfnamefont {N.-H.}\ \bibnamefont {Kim}},
  \bibinfo {author} {\bibfnamefont {S.}~\bibnamefont {Lee}}, \bibinfo {author}
  {\bibfnamefont {R.}~\bibnamefont {Lavrijsen}}, \bibinfo {author}
  {\bibfnamefont {A.}~\bibnamefont {Solignac}}, \bibinfo {author}
  {\bibfnamefont {Y.}~\bibnamefont {Yin}}, \bibinfo {author} {\bibfnamefont
  {D.-S.}\ \bibnamefont {Han}}, \bibinfo {author} {\bibfnamefont {N.~J.~J.}\
  \bibnamefont {van Hoof}}, \bibinfo {author} {\bibfnamefont {H.~J.~M.}\
  \bibnamefont {Swagten}}, \bibinfo {author} {\bibfnamefont {B.}~\bibnamefont
  {Koopmans}}, \bibinfo {author} {\bibfnamefont {J.-S.}\ \bibnamefont {Kim}}, \
  and\ \bibinfo {author} {\bibfnamefont {C.-Y.}\ \bibnamefont {You}},\ }\href
  {http://dx.doi.org/10.1038/ncomms8635} {\bibfield  {journal} {\bibinfo
  {journal} {Nat. Commun.}\ }\textbf {\bibinfo {volume} {6}},\ \bibinfo {pages}
  {1} (\bibinfo {year} {2015})}\BibitemShut {NoStop}%
\bibitem [{\citenamefont {Chauve}\ \emph {et~al.}(2000)\citenamefont {Chauve},
  \citenamefont {Giamarchi},\ and\ \citenamefont {Le~Doussal}}]{Chauve:2000}%
  \BibitemOpen
  \bibfield  {author} {\bibinfo {author} {\bibfnamefont {P.}~\bibnamefont
  {Chauve}}, \bibinfo {author} {\bibfnamefont {T.}~\bibnamefont {Giamarchi}}, \
  and\ \bibinfo {author} {\bibfnamefont {P.}~\bibnamefont {Le~Doussal}},\
  }\href {\doibase 10.1103/PhysRevB.62.6241} {\bibfield  {journal} {\bibinfo
  {journal} {Phys. Rev. B}\ }\textbf {\bibinfo {volume} {62}},\ \bibinfo
  {pages} {6241} (\bibinfo {year} {2000})}\BibitemShut {NoStop}%
\bibitem [{\citenamefont {Thiaville}\ \emph {et~al.}(2012)\citenamefont
  {Thiaville}, \citenamefont {Rohart}, \citenamefont {Ju\'e}, \citenamefont
  {Cros},\ and\ \citenamefont {Fert}}]{Thiaville:2012}%
  \BibitemOpen
  \bibfield  {author} {\bibinfo {author} {\bibfnamefont {A.}~\bibnamefont
  {Thiaville}}, \bibinfo {author} {\bibfnamefont {S.}~\bibnamefont {Rohart}},
  \bibinfo {author} {\bibfnamefont {E.}~\bibnamefont {Ju\'e}}, \bibinfo
  {author} {\bibfnamefont {V.}~\bibnamefont {Cros}}, \ and\ \bibinfo {author}
  {\bibfnamefont {A.}~\bibnamefont {Fert}},\ }\href
  {http://stacks.iop.org/0295-5075/100/i=5/a=57002} {\bibfield  {journal}
  {\bibinfo  {journal} {Europhys. Lett.}\ }\textbf {\bibinfo {volume} {100}},\
  \bibinfo {pages} {57002} (\bibinfo {year} {2012})}\BibitemShut {NoStop}%
\bibitem [{\citenamefont {Vansteenkiste}\ \emph {et~al.}(2014)\citenamefont
  {Vansteenkiste}, \citenamefont {Leliaert}, \citenamefont {Dvornik},
  \citenamefont {Helsen}, \citenamefont {Garcia-Sanchez},\ and\ \citenamefont
  {Van~Waeyenberge}}]{Vansteenkiste:2014}%
  \BibitemOpen
  \bibfield  {author} {\bibinfo {author} {\bibfnamefont {A.}~\bibnamefont
  {Vansteenkiste}}, \bibinfo {author} {\bibfnamefont {J.}~\bibnamefont
  {Leliaert}}, \bibinfo {author} {\bibfnamefont {M.}~\bibnamefont {Dvornik}},
  \bibinfo {author} {\bibfnamefont {M.}~\bibnamefont {Helsen}}, \bibinfo
  {author} {\bibfnamefont {F.}~\bibnamefont {Garcia-Sanchez}}, \ and\ \bibinfo
  {author} {\bibfnamefont {B.}~\bibnamefont {Van~Waeyenberge}},\ }\href
  {\doibase http://dx.doi.org/10.1063/1.4899186} {\bibfield  {journal}
  {\bibinfo  {journal} {AIP Adv.}\ }\textbf {\bibinfo {volume} {4}},\ \bibinfo
  {eid} {107133} (\bibinfo {year} {2014})}\BibitemShut {NoStop}%
\bibitem [{\citenamefont {Burrowes}\ \emph {et~al.}(2013)\citenamefont
  {Burrowes}, \citenamefont {Vernier}, \citenamefont {Adam}, \citenamefont
  {Herrera~Diez}, \citenamefont {Garcia}, \citenamefont {Barisic},
  \citenamefont {Agnus}, \citenamefont {Eimer}, \citenamefont {Kim},
  \citenamefont {Devolder}, \citenamefont {Lamperti}, \citenamefont {Mantovan},
  \citenamefont {Ockert}, \citenamefont {Fullerton},\ and\ \citenamefont
  {Ravelosona}}]{Burrowes:2013}%
  \BibitemOpen
  \bibfield  {author} {\bibinfo {author} {\bibfnamefont {C.}~\bibnamefont
  {Burrowes}}, \bibinfo {author} {\bibfnamefont {N.}~\bibnamefont {Vernier}},
  \bibinfo {author} {\bibfnamefont {J.-P.}\ \bibnamefont {Adam}}, \bibinfo
  {author} {\bibfnamefont {L.}~\bibnamefont {Herrera~Diez}}, \bibinfo {author}
  {\bibfnamefont {K.}~\bibnamefont {Garcia}}, \bibinfo {author} {\bibfnamefont
  {I.}~\bibnamefont {Barisic}}, \bibinfo {author} {\bibfnamefont
  {G.}~\bibnamefont {Agnus}}, \bibinfo {author} {\bibfnamefont
  {S.}~\bibnamefont {Eimer}}, \bibinfo {author} {\bibfnamefont {J.-V.}\
  \bibnamefont {Kim}}, \bibinfo {author} {\bibfnamefont {T.}~\bibnamefont
  {Devolder}}, \bibinfo {author} {\bibfnamefont {A.}~\bibnamefont {Lamperti}},
  \bibinfo {author} {\bibfnamefont {R.}~\bibnamefont {Mantovan}}, \bibinfo
  {author} {\bibfnamefont {B.}~\bibnamefont {Ockert}}, \bibinfo {author}
  {\bibfnamefont {E.~E.}\ \bibnamefont {Fullerton}}, \ and\ \bibinfo {author}
  {\bibfnamefont {D.}~\bibnamefont {Ravelosona}},\ }\href {\doibase
  http://dx.doi.org/10.1063/1.4826439} {\bibfield  {journal} {\bibinfo
  {journal} {Appl. Phys. Lett.}\ }\textbf {\bibinfo {volume} {103}},\ \bibinfo
  {eid} {182401} (\bibinfo {year} {2013})}\BibitemShut {NoStop}%
\bibitem [{\citenamefont {Yamada}\ \emph {et~al.}(2011)\citenamefont {Yamada},
  \citenamefont {Jamet}, \citenamefont {Nakatani}, \citenamefont {Mougin},
  \citenamefont {Thiaville}, \citenamefont {Ono},\ and\ \citenamefont
  {Ferr{\'e}}}]{Yamada:2011}%
  \BibitemOpen
  \bibfield  {author} {\bibinfo {author} {\bibfnamefont {K.}~\bibnamefont
  {Yamada}}, \bibinfo {author} {\bibfnamefont {J.-P.}\ \bibnamefont {Jamet}},
  \bibinfo {author} {\bibfnamefont {Y.}~\bibnamefont {Nakatani}}, \bibinfo
  {author} {\bibfnamefont {A.}~\bibnamefont {Mougin}}, \bibinfo {author}
  {\bibfnamefont {A.}~\bibnamefont {Thiaville}}, \bibinfo {author}
  {\bibfnamefont {T.}~\bibnamefont {Ono}}, \ and\ \bibinfo {author}
  {\bibfnamefont {J.}~\bibnamefont {Ferr{\'e}}},\ }\href@noop {} {\bibfield
  {journal} {\bibinfo  {journal} {Appl. Phys. Express}\ }\textbf {\bibinfo
  {volume} {4}},\ \bibinfo {pages} {113001} (\bibinfo {year}
  {2011})}\BibitemShut {NoStop}%
\bibitem [{\citenamefont {Weindler}\ \emph {et~al.}(2014)\citenamefont
  {Weindler}, \citenamefont {Bauer}, \citenamefont {Islinger}, \citenamefont
  {Boehm}, \citenamefont {Chauleau},\ and\ \citenamefont
  {Back}}]{Weindler2014}%
  \BibitemOpen
  \bibfield  {author} {\bibinfo {author} {\bibfnamefont {T.}~\bibnamefont
  {Weindler}}, \bibinfo {author} {\bibfnamefont {H.~G.}\ \bibnamefont {Bauer}},
  \bibinfo {author} {\bibfnamefont {R.}~\bibnamefont {Islinger}}, \bibinfo
  {author} {\bibfnamefont {B.}~\bibnamefont {Boehm}}, \bibinfo {author}
  {\bibfnamefont {J.-Y.}\ \bibnamefont {Chauleau}}, \ and\ \bibinfo {author}
  {\bibfnamefont {C.~H.}\ \bibnamefont {Back}},\ }\href {\doibase
  10.1103/PhysRevLett.113.237204} {\bibfield  {journal} {\bibinfo  {journal}
  {Phys. Rev. Lett.}\ }\textbf {\bibinfo {volume} {113}},\ \bibinfo {pages}
  {237204} (\bibinfo {year} {2014})}\BibitemShut {NoStop}%
\bibitem [{\citenamefont {Kim}(2015)}]{Kim:2015}%
  \BibitemOpen
  \bibfield  {author} {\bibinfo {author} {\bibfnamefont {J.-V.}\ \bibnamefont
  {Kim}},\ }\href@noop {} {\bibfield  {journal} {\bibinfo  {journal} {Phys.
  Rev. B}\ }\textbf {\bibinfo {volume} {92}},\ \bibinfo {pages} {014418}
  (\bibinfo {year} {2015})}\BibitemShut {NoStop}%
\bibitem [{\citenamefont {Gross}\ \emph {et~al.}()\citenamefont {Gross},
  \citenamefont {Martinez}, \citenamefont {Tetienne}, \citenamefont {Hingant},
  \citenamefont {Roch}, \citenamefont {Garcia}, \citenamefont {Soucaille},
  \citenamefont {Adam}, \citenamefont {Kim}, \citenamefont {Rohart},
  \citenamefont {Thiaville}, \citenamefont {Torrejon}, \citenamefont
  {Hayashi},\ and\ \citenamefont {Jacques}}]{Gross_NV}%
  \BibitemOpen
  \bibfield  {author} {\bibinfo {author} {\bibfnamefont {I.}~\bibnamefont
  {Gross}}, \bibinfo {author} {\bibfnamefont {L.~J.}\ \bibnamefont {Martinez}},
  \bibinfo {author} {\bibfnamefont {J.-P.}\ \bibnamefont {Tetienne}}, \bibinfo
  {author} {\bibfnamefont {T.}~\bibnamefont {Hingant}}, \bibinfo {author}
  {\bibfnamefont {J.-F.}\ \bibnamefont {Roch}}, \bibinfo {author}
  {\bibfnamefont {K.}~\bibnamefont {Garcia}}, \bibinfo {author} {\bibfnamefont
  {R.}~\bibnamefont {Soucaille}}, \bibinfo {author} {\bibfnamefont {J.-P.}\
  \bibnamefont {Adam}}, \bibinfo {author} {\bibfnamefont {J.-V.}\ \bibnamefont
  {Kim}}, \bibinfo {author} {\bibfnamefont {S.}~\bibnamefont {Rohart}},
  \bibinfo {author} {\bibfnamefont {A.}~\bibnamefont {Thiaville}}, \bibinfo
  {author} {\bibfnamefont {J.}~\bibnamefont {Torrejon}}, \bibinfo {author}
  {\bibfnamefont {M.}~\bibnamefont {Hayashi}}, \ and\ \bibinfo {author}
  {\bibfnamefont {V.}~\bibnamefont {Jacques}},\ }\href@noop {} {\bibinfo
  {journal} {unpublished}\ }\BibitemShut {NoStop}%
\end{thebibliography}
\end{document}